\documentclass[journal]{IEEEtran}

\ifCLASSINFOpdf
\else
\fi
\usepackage{amsmath}
\usepackage{comment}
\usepackage{cancel}
\usepackage{graphicx}
\usepackage{amssymb}
\usepackage{xcolor}
\usepackage{cite}
\usepackage{physics}
\makeatletter
\newcommand*{\rom}[1]{\expandafter\@slowromancap\romannumeral #1@}
\makeatother

\def\Xint#1{\mathchoice
   {\XXint\displaystyle\textstyle{#1}}%
   {\XXint\textstyle\scriptstyle{#1}}%
   {\XXint\scriptstyle\scriptscriptstyle{#1}}%
   {\XXint\scriptscriptstyle\scriptscriptstyle{#1}}%
   \!\int}
\def\XXint#1#2#3{{\setbox0=\hbox{$#1{#2#3}{\int}$}
     \vcenter{\hbox{$#2#3$}}\kern-.5\wd0}}

\def\dashint{\Xint-}

\begin{document}
%
\title{Analytic Preconditioners for Decoupled Potential Integral Equations and Wideband Analysis of Scattering from PEC Objects}
%
%
%

\author{J. A. Hawkins,~\IEEEmembership{Student Member,~IEEE,}
        L. Baumann,~\IEEEmembership{Student Member,~IEEE,}
        H. M. Aktulga,~\IEEEmembership{Member,~IEEE,} \\
        D. Dault, ~\IEEEmembership{Senior Member,~IEEE,}
        B. Shanker,~\IEEEmembership{Fellow,~IEEE}
\thanks{J. A. Hawkins and L. Baumann are with the Department of Electrical and Computer Engineering, Michigan State University, East Lansing, MI 48823 USA}
\thanks{H. M. Aktulga is with the Department of Computer Science, Michigan State University, East Lansing, MI 48823 USA }
\thanks{D. Dault is with the Air Force Research Laboratory, Wright-Patterson AFB, OH 45433 USA}
\thanks{B. Shanker is with the Department of Electrical and Computer Engineering, The Ohio State University, Columbus, OH 43210 USA}}

\markboth{IEEE Transactions on Antennas and Propagation Journal,~Vol.~xx, No.~xx, xx 2022}%
{Shell \MakeLowercase{\textit{et al.}}: Analytic preconditioners for decoupled potential integral equations and wideband analysis of scattering from PEC objects}

\maketitle

\begin{abstract}

Many integral equations used to analyze scattering, such as the standard combined field integral equation (CFIE), are not well-conditioned for a wide range of frequencies and multi-scale geometries. There has been significant effort to alleviate this problem. A more recent one is using a set of decoupled potential integral equations (DPIE). These equations have been shown to be robust at low frequencies and immune to topology breakdown. But they  mimic the ill-conditioning behavior of CFIE at high frequencies. This paper addresses this deficiency through new Calder\'{o}n-type identities derived from the Vector Potential Integral Equation (VPIE). We construct novel analytic preconditioners for the vector potential integral equation (VPIE) and scalar potential integral equation (SPIE) constrained to perfect electric conductors (PEC). These new formulations are wide-band well-conditioned and converge rapidly for multi-scale geometries. This is demonstrated though a number of examples that use analytic and piecewise basis sets. 

\end{abstract}


%
\IEEEpeerreviewmaketitle

\section{Introduction}\label{sec:intro}
%
%
%
%

\IEEEPARstart{R}{esearchers} have developed various boundary integral equation approaches to predict scattered magnetic and electric fields from PEC objects of arbitrary shape. When applicable, the computation time of this approach is less than that of other approaches such as the Finite Element Method (FEM)\cite{Peterson1998CEM}, and the Silver-M\"{u}ller radiation condition is automatically enforced. One classic perfect electric conductor (PEC) boundary integral equation is the combined field integral equation (CFIE)\cite{ref:Harrington1989CFIE}\cite{ref:Harrington1978BOR}, where the electric field integral equation (EFIE) and magnetic field integral equation (MFIE) are combined to fashion a provably unique formulation avoiding the spurious modes latent in other non-unique formulations\cite{Peterson1998CEM}.

The EFIE and MFIE components of the CFIE formulation suffer from a variety of issues like low-frequency breakdown\cite{Yan2010EFIE}\cite{ref:Zhao2000LF}\cite{ref:Zhang2003MFIE}\cite{ref:Qian2008LF}, catastrophic cancellation\cite{ref:Kress1981LF}, dense mesh breakdown\cite{ref:Valdes2011Dense}, static nullspaces or topology breakdown\cite{ref:Cools2009Null}, and a poor approximation of the identity operator in the MFIE with RWG testing and basis functions\cite{ref:Yan2011MFIE}. These issues have been addressed in various ways. Buffa-Christiansen testing sets better approximate the identity operator of the MFIE when the basis set is composed of RWG functions \cite{ref:Cools2009MFIE}. These functions have also been used in conjunction with the Calder\'{o}n identities to precondition the ill-conditioned EFIE operator \cite{ref:Cools2009Calderon}. Other suggested basis sets to alleviate breakdown are the so-called loop-star and loop-tree functions\cite{ref:Glisson1981Loop}\cite{ref:Glisson1995Loop} or the related basis-free quasi-Helmholtz projection matrices\cite{ref:Andruilli2013LS}, subdivision surfaces \cite{ref:Fu2017Subdivision}, and manifold harmonics\cite{ref:Alsnayyan2021Manifold}. Yet another option is solving for current and charge densities in the current and charge integral equation (CCIE)\cite{ref:Taskinen2006Charge}.

The focus of this paper is to use decoupled potentials instead of fields. The resulting equations are the scalar potential integral equation (SPIE) and the vector potential integral equation (VPIE) that are solved independently. For PEC objects analyzed in the frequency domain, \cite{Li2019DPIE} suggested various combinations of the potential integral equations to construct well-conditioned formulations at low frequency. Reference \cite{Eris2022VPIE} suggested a unique combined potential formulation for dense discretizations. The low frequency behavior of the VPIE was further analyzed for PEC in \cite{Chen2022VPIE}. For dielectric objects analyzed in the frequency domain, \cite{Li2019DPIE}\cite{vico2016DPIE} have suggested well-conditioned potential formulations at low frequency. A potential integral formulation for solving lossy conductors is detailed in \cite{Triverio2022VPIE} as well. The time-domain variant of these integral equations has been analyzed in \cite{ref:RothDPIETD2021}. Specifically, the decoupled potential integral equation (DPIE) approach has several niceties: no low-frequency breakdown, no dense mesh breakdown, no topological low-frequency breakdown, and well-conditioned dielectric and PEC formulations at low to medium range frequencies \cite{vico2016DPIE}\cite{Li2019DPIE}. Recently, the DPIE has been implemented on arbitrary dielectric objects \cite{ref:Baumann2022DPIE} with pulse and RWG functions, and the results therein demonstrate low singular value conditioning and a low iteration count to converge. However, fast convergence with iterative solvers has yet to be demonstrated in the high frequency region for arbitrary objects using the DPIE.

This paper extends the DPIE property of well-conditioned to the high frequency region for arbitrary PEC objects by constructing a new formulation through novel Calder\'{o}n-type identities. Several salient contributions are made along the way. One, new integral equation identities are derived in a similar manner to the derivation of the Calder\'{o}n identities in \cite{Hsiao97calderon}, and the process is applicable to other scattering problems whose forward matrices are also projectors. Two, we show the new Calder\'{o}n-type identities are effective in preconditioning a combined VPIE and SPIE formulation while maintaining uniqueness. Three, the new combined formulations map a div-conforming space to a div-conforming space and a curl-conforming space to a curl-conforming space. And finally, the new formulation quickly converges in practice for tessellations of multi-scale geometries using RWG and hat testing and basis functions. 

This paper is organized as follows: the problem is stated in Section \rom{2}; the potential integral equations are given in Section \rom{3}; the Calder\'{o}n-type identities are derived in Section \rom{4}; the VPIE and SPIE formulations for the PEC problem are derived in Section \rom{5}; novel VPIE and SPIE formulations are developed in Section \rom{6}; the implementation is discussed in Section \rom{7}; and results are presented in Section \rom{8}.

\section{Problem Statement}

Consider a PEC closed object of a simply connected region $D_{-}$ immersed in a homogeneous background $D_{+}$. The object's boundary is a two-dimensional smooth manifold $S = \partial D$ embedded in $\mathbb{R}^{3}$ with unique normal $\hat{\textbf{n}}(\textbf{r})$ pointing from $\partial\Omega$ into $D_{+}$. It is assumed that a plane wave characterized by $\left \{\boldsymbol{\kappa},\textbf{E}^{i}(\textbf{r}), \textbf{H}^{i}(\textbf{r})\right \}$ is incident on the object. Here, $\boldsymbol{\kappa} = \kappa \hat{\boldsymbol{\kappa}}$ is the wave vector, and $\kappa = \omega\sqrt{\epsilon\mu}$ being the wave number. Our objective is to determine scattered fields in $D_+$. In what follows, all quantities associated with $D_\pm$ will use the appropriate subscript. As is usually  done, $\left\{ \textbf{E}_+(\textbf{r}), \textbf{H}_+(\textbf{r}) \right \}$ = $\left \{ \textbf{E}^{s}(\textbf{r}), \textbf{H}^{s}(\textbf{r}) \right \}$, which satisfies the Silver-Muller radiation condition,  and $\left\{ \textbf{E}_-(\textbf{r}), \textbf{H}_-(\textbf{r}) \right \}$ = $\left \{ \textbf{E}^{i}(\textbf{r}), \textbf{H}^{i}(\textbf{r}) \right \}$ \cite{Hsiao97calderon}. As opposed to the usual approach of analysis using equivalent currents and fields, we will formulate this problem in terms of decoupled potentials. Note, the time dependence $e^{j\omega t}$ is assumed and suppressed throughout. 

\section{Potential Integral Representations}

DPIE takes a different approach to solve for scattering from perfectly conducting objects. As alluded to in the Section \ref{sec:intro}, we define decoupled scalar electric and vector magnetic potentials for regions both exterior and interior to $S$ (that is, in $D^\pm$) as $\textbf{A}_{\pm}(\textbf{r})$ and $\phi_{\pm}(\textbf{r})$. Using 
\begin{equation}
    \begin{split}
        \mathcal{u} (\textbf{r}) = & 1~~~~\vb {r} \in D_{+} \\
        \mathcal{u} (\textbf{r}) = & 0~~~~\vb{r}\in D_{-} \\
    \end{split}
\end{equation}
and the Green's identity, the magnetic vector potential can be written as 
\begin{equation}
    \begin{split}
        \mathcal{u} (\textbf{r})\textbf{A}_{+}(\textbf{r}) + [\mathcal{u} (\textbf{r}) - 1]\textbf{A}_{-}(\textbf{r}) & = \\
        -\int_{S} G(\textbf{r},\textbf{r}')~\hat{\textbf{n}}'\times\textbf{a}_{\pm}(\textbf{r}')dS' & +
        \nabla\times\int_{S} G(\textbf{r},\textbf{r}')~\textbf{b}_{\pm}(\textbf{r}')dS' \\
          - \nabla\int_{S} G(\textbf{r},\textbf{r}')~ \gamma_{\pm} (\textbf{r}')dS' &- 
         \int_{S} G(\textbf{r},\textbf{r}')~\hat{\textbf{n}}'\sigma_{\pm} (\textbf{r}')dS'  \\
    \end{split}
\end{equation}
where the source terms are defined as, 
\begin{equation}
\begin{split}
    \textbf{a}(\textbf{r}) = &\hat{\mathbf{n}}\times\hat{\mathbf{n}}\times\nabla\times\textbf{A}(\textbf{r}) \\
    \textbf{b}(\textbf{r}) = & \hat{\mathbf{n}}\times\textbf{A}(\textbf{r}) \\
    \gamma(\textbf{r}) = & \hat{\mathbf{n}}\cdot\textbf{A}(\textbf{r}) \\
    \sigma(\textbf{r}) = & \nabla\cdot\textbf{A}(\textbf{r}). \\
\end{split}
\end{equation}
In a a similar manner, the scalar potential can be written as 
\begin{equation}
    \begin{split}
        &\mathcal{u} (\textbf{r})\phi_{+}(\textbf{r}) + [\mathcal{u} (\textbf{r}) - 1]\phi_{-}(\textbf{r})= \\ &-\nabla\cdot\int_{S} G(\textbf{r},\textbf{r}')~\hat{\textbf{n}}'\alpha_{\pm}(\textbf{r}')dS' -
        \int_{S}G(\textbf{r},\textbf{r}')\beta_{\pm}(\textbf{r}')dS' \\
    \end{split}
\end{equation}
where the source terms are $\alpha (\textbf{r}) = \phi(\textbf{r})$ and $\beta(\textbf{r}) = \hat{\mathbf{n}}\cdot\nabla\phi(\textbf{r})$ and,
\begin{equation}
    G(\textbf{r},\textbf{r}') = \frac{e^{-j\kappa(\textbf{r} - \textbf{r}')}}{4\pi(\textbf{r} - \textbf{r}')}
\end{equation}
is the free space Green's function. We note that  $\gamma (\textbf{r})$ and $\mathbf{b}(\textbf{r})$ are H\"{o}lder continuous and $\sigma (\textbf{r})$ and $\hat{\mathbf{n}} \cross \mathbf{a}(\textbf{r})$ are continuous. In the context of scattering, we associate quantities denoted by $\pm$ with scattered fields and incident fields, respectively. That is $\textbf{A}_{-}(\textbf{r})\rightarrow{\textbf{A}^{i}(\textbf{r})}$ and $\textbf{A}_{+}(\textbf{r})\rightarrow{\textbf{A}^{s}(\textbf{r})}$. Using the classical jump conditions, adding the two forms $\textbf{A}_{\pm}$, and denoting $\textbf{A}(\textbf{r}) = \textbf{A}^i(\textbf{r}) + \textbf{A}^s(\textbf{r})$, we have 
\begin{equation}
\label{eq:VPIE}
    \begin{split}
        \frac{1}{2}\textbf{A}(\textbf{r}) - \textbf{A}^i(\textbf{r}) = &
        \dashint_{S}\bigg[ -G(\textbf{r},\textbf{r}')~\hat{\textbf{n}}'\times\textbf{a}(\textbf{r}') + \\ &G(\textbf{r},\textbf{r}')\nabla\times\textbf{b}(\textbf{r}') - \nabla G(\textbf{r},\textbf{r}')~ \gamma (\textbf{r}') - \\ & G(\textbf{r},\textbf{r}')~\hat{\textbf{n}}'\sigma (\textbf{r}')\bigg]dS'.
    \end{split}
\end{equation}
It can be shown that the quantities associated with the scalar potential, namely $\hat{\textbf{n}}\alpha(\textbf{r})$ and $\beta(\textbf{r})$, are H\"older continuous and continuous, respectively.  As before, we associate the scattered scalar potential with scalar potential in the exterior region and the incident scalar potential with the interior region ( $\phi_{-}(\textbf{r})\rightarrow{\phi^{i}(\textbf{r})}$ and $\phi_{+}(\textbf{r})\rightarrow{\phi^{s}(\textbf{r})}$). Using classical jump conditions, adding the two forms $\phi_{\pm}$ and denoting $\phi(\textbf{r}) = \phi^i(\textbf{r}) + \phi^s(\textbf{r})$, we obtain
\begin{equation}
\label{eq:SPIE}
    \begin{split}
        \frac{1}{2}\phi(\textbf{r}) - \phi^i(\textbf{r}) =  
        \frac{1}{4\pi}\dashint_{S}\bigg[-\nabla\mathcal{G}(\textbf{r},\textbf{r}')\cdot\hat{\textbf{n}}'\alpha (\textbf{r}') - \\ \mathcal{G}(\textbf{r},\textbf{r}')~\beta (\textbf{r}')\bigg]dS'. \\
    \end{split}
\end{equation}
While these equations hold in general, they simplify considerably when $S$ is a PEC. $\mathbf{b}(\vb{r})$, $\sigma(\vb{r})$ and $\alpha(\vb{r})$ are zero  \cite{Li2019DPIE}. This is a point that we save for later in the paper. 

With \eqref{eq:VPIE} and \eqref{eq:SPIE} and the decoupled potential boundary conditions discussed in \cite{Li2019DPIE}, one can obtain a set of integral equations for both the scalar and vector potentials through the boundary conditions. As reported in \cite{Li2019DPIE} and \cite{ref:Baumann2022DPIE}, these equations that can be succinctly written as
\begin{subequations}
\begin{equation}
\label{eq:SPIEEqns}
    \left ( \mathcal{I} - \mathcal{Z}^{SPIE} \right ) \begin{pmatrix}
    \alpha \\
    \beta \\
\end{pmatrix} = \begin{pmatrix}
    \alpha^i \\
    \beta^i \\
\end{pmatrix}
\end{equation} 
and
\begin{equation}
\label{eq:VPIEEqns}
    \left ( \mathcal{I} - \mathcal{Z}^{VPIE}\right ) \begin{pmatrix}
    \textbf{a} \\
    \textbf{b} \\
    \gamma \\
    \sigma \\
    \end{pmatrix} = 
    \begin{pmatrix}
    \textbf{a}^i \\
    \textbf{b}^i \\
    \gamma^i \\
    \sigma^i \\
    \end{pmatrix}
\end{equation}
where
\begin{equation}
\begin{split}\
   \mathcal{Z}^{SPIE} = & \begin{pmatrix}\label{eq:SPIEsystem}
    \mathcal{D} & -\mathcal{S} \\
    \mathcal{N} & -\mathcal{D}' \\
    \end{pmatrix} \\
\end{split}
\end{equation}
and 
\begin{equation}\label{eq:VPIEsystem}
    \mathcal{Z}^{VPIE} = \begin{pmatrix}
        \mathcal{K}^{'t} & \kappa^2\mathcal{L}^{t} & 0 & -\mathcal{Q}^{(1)} \\
        -\mathcal{J}^{(2)} & \mathcal{K} & -\mathcal{P}^{(2)} & -\mathcal{Q}^{(2)} \\
        -\mathcal{J}^{(3)} & \mathcal{M}^{(3)} & -\mathcal{D}' & -\mathcal{Q}^{(3)} \\
        -\mathcal{J}^{(4)} & 0 & \kappa^{2}\mathcal{S} & \mathcal{D} \\
    \end{pmatrix}.
\end{equation}
\end{subequations}
The terms \emph{SPIE} and \emph{VPIE} denote the scalar and vector potential integral equations, respectively, and the operators are defined in Appendix A. Note, we have not separated out the principal value components in the diagonal elements of $\mathcal{Z}^{SPIE/VPIE}$.

\section{Novel Calder\'{o}n-Type Identities}\label{sec:CalderonIdentities}

Next, we derive Calder\'{o}n-type identities for both the SPIE and VPIE. To do so, we follow a similar procedure outlined in  \cite{Hsiao97calderon} but stress the perspective of projectors; namely, we use the projector and complement projector properties of $\mathcal{Z}^{SPIE/VPIE}$ and $\mathcal{I} - \mathcal{Z}^{SPIE/VPIE}$. Indeed, Calder\'{o}n-type identities may be derived for any scattering problem provided the forward matrix is a projector.

In the absence of incident potentials in both \eqref{eq:SPIEEqns} and \eqref{eq:VPIEEqns}, one can show that 
\begin{subequations}
 \begin{equation}
\begin{split}
\begin{pmatrix}
0 \\
0 \\
\end{pmatrix} = (\mathcal{I} - \mathcal{Z}^{SPIE})\mathcal{Z}^{SPIE}\begin{pmatrix}
\alpha \\
\beta \\
\end{pmatrix}
\end{split}
\end{equation}
and 
\begin{equation}
\begin{split}
\begin{pmatrix}
0 \\
0 \\
0 \\
0 \\
\end{pmatrix} = & (\mathcal{I} - \mathcal{Z}^{VPIE})\mathcal{Z}^{VPIE}\begin{pmatrix}
    \textbf{a} \\
    \textbf{b} \\
    \mathbf{\gamma} \\
    \mathbf{\sigma} \\
\end{pmatrix}.
\end{split}
\end{equation}
\end{subequations}
Note, $(\mathcal{I} - \mathcal{Z}^{SPIE/VPIE})$ and $\mathcal{Z}^{SPIE/VPIE}$ commute. As the sources are arbitrary, it follows that the cascaded operators have to be zero. This results in the well-known identities
\begin{subequations}
\label{eq:ScalarCalderonIdentities}
\begin{align}
    (\mathcal{I} - \mathcal{D})\mathcal{D} + \mathcal{S}\mathcal{N}  = & ~0 \\
    -(\mathcal{I - \mathcal{D}})\mathcal{S} - \mathcal{S}\mathcal{D}' = & ~0 \\
    -\mathcal{N}\mathcal{D} + (\mathcal{I} + \mathcal{D}')\mathcal{N}  = & ~0 \\
    \mathcal{N}\mathcal{S} - (\mathcal{I} + \mathcal{D}')\mathcal{D}' = & ~0 \label{eq:NS_identity}
\end{align}
\end{subequations}
for the SPIE; note these have been previously derived in  \cite{Nedelec2001EM} for scalar potentials and acoustics. As before, these operator equations are meant to be read from right to left. For instance, $\mathcal{SN}\alpha = \mathcal{S} \left ( \mathcal{N} \alpha \right )$. Following a similar procedure for the VPIE system, we have
\begin{subequations}\label{eq:VectorCalderonIdentities}
\begin{align}
    (\mathcal{I} - \mathcal{K}^{'t})\mathcal{K}^{'t} + \kappa^2\mathcal{L}^{t}\mathcal{J}^{(2)} - \mathcal{Q}^{(1)}\mathcal{J}^{(4)} = & ~0 \label{NewCalderon_a}\\ 
    \mathcal{J}^{(2)}\mathcal{K}^{'t} - (\mathcal{I} - \mathcal{K})\mathcal{J}^{(2)} - \mathcal{P}^{(2)}\mathcal{J}^{(3)} - \mathcal{Q}^{(2)}\mathcal{J}^{(4)} = & ~0 \label{NewCalderon_b}\\
    \mathcal{J}^{(3)}\mathcal{K}^{'t} + \mathcal{M}^{(3)}\mathcal{J}^{(2)} - (\mathcal{I} + \mathcal{D}')\mathcal{J}^{(3)} - \mathcal{Q}^{(3)}\mathcal{J}^{(4)} = & ~0 \label{NewCalderon_c}\\ 
    \mathcal{J}^{(4)}\mathcal{K}^{'t} + \kappa^2\mathcal{S}\mathcal{J}^{(3)} - (\mathcal{I} - \mathcal{D})\mathcal{J}^{(4)} = & ~0 \label{NewCalderon_d}\\
    (\mathcal{I} - \mathcal{K}^{'t})\mathcal{L}^{t} - \mathcal{L}^{t}\mathcal{K} = &  ~0 \label{NewCalderon_e}\\ 
    \mathcal{J}^{(2)}\kappa^2\mathcal{L}^{t} + (\mathcal{I} - \mathcal{K})\mathcal{K} + \mathcal{P}^{(2)}\mathcal{M}^{(3)} = & ~0 \label{NewCalderon_f}\\ 
    \mathcal{J}^{(3)}\kappa^2\mathcal{L}^{t} - \mathcal{M}^{(3)}\mathcal{K} + (\mathcal{I} + \mathcal{D}')\mathcal{M}^{(3)} = & ~0 \label{NewCalderon_g}\\ 
    \mathcal{J}^{(4)}\mathcal{L}^{t} - \mathcal{S}\mathcal{M}^{(3)} = & ~0 \label{NewCalderon_h}\\ 
    \mathcal{L}^{t}\mathcal{P}^{(2)} + \mathcal{Q}^{(1)}\mathcal{S} = & ~0 \label{NewCalderon_i}\\ 
    -(\mathcal{I} - \mathcal{K})\mathcal{P}^{(2)} - \mathcal{P}^{(2)}\mathcal{D}' + \mathcal{Q}^{(2)}\kappa^2\mathcal{S} = & ~0 \label{NewCalderon_j}\\ 
    \mathcal{M}^{(3)}\mathcal{P}^{(2)} - (\mathcal{I} + \mathcal{D}')\mathcal{D}' + \mathcal{Q}^{(3)}\kappa^2\mathcal{S} = & ~0 \label{NewCalderon_k} \\ 
    \mathcal{S}\mathcal{D}' + (\mathcal{I} - \mathcal{D})\mathcal{S} = & ~0 \label{NewCalderon_l}\\ 
    -(\mathcal{I} - \mathcal{K}^{'t})\mathcal{Q}^{(1)} + \kappa^2\mathcal{L}^{t}\mathcal{Q}^{(2)} + \mathcal{Q}^{(1)}\mathcal{D} = & ~0 \label{NewCalderon_m} \\ 
    -\mathcal{J}^{(2)}\mathcal{Q}^{(1)} - (\mathcal{I} - \mathcal{K})\mathcal{Q}^{(2)} - \mathcal{P}^{(2)}\mathcal{Q}^{(3)} +\mathcal{Q}^{(2)}\mathcal{D} = & ~0 \label{NewCalderon_n} \\ 
    -\mathcal{J}^{(3)}\mathcal{Q}^{(1)} + \mathcal{M}^{(3)}\mathcal{Q}^{(2)} - (\mathcal{I} + \mathcal{D}')\mathcal{Q}^{(3)} + \mathcal{Q}^{(3)}\mathcal{D} = & ~0 \label{NewCalderon_o} \\ 
    -\mathcal{J}^{(4)}\mathcal{Q}^{(1)} + \kappa^2\mathcal{S}\mathcal{Q}^{(3)} + (\mathcal{I} - \mathcal{D})\mathcal{D} = & ~0 \label{NewCalderon_p}.
\end{align}
\end{subequations}
We will leverage these identities to precondition VPIE and SPIE systems as applied to the analysis of scattering from perfect electrically conducting objects. Parenthetically, we note the following: Given the similarity of operators to those for the Decoupled Field Integral Equation (DFIE)\cite{ref:Vico2018DFIE}, the same approach can be used for these as well, and we will report the results in a future paper.

Next, we present the reduction necessary for the analysis of scattering from PEC objects. Note that the boundary conditions should enforce that the tangential component of the total electric field is zero. For potentials, this is tantamount to setting $\alpha (\vb{r}) = 0 = \sigma(\vb{r})$ and $\vb{b}(\vb{r}) = \underline{0}$. These result in equations of the form
\begin{subequations}
    \begin{equation}\label{eq:SPIE_pec}
    \begin{pmatrix}
    \alpha^i \\
    \beta^i \\
    \end{pmatrix}
     = \begin{pmatrix}
        \mathcal{S} \\
        \mathcal{I} + \mathcal{D}' \\
    \end{pmatrix}
    \beta \\
\end{equation}
\begin{equation}\label{eq:VPIE_pec}
    \begin{pmatrix}
    \textbf{a}^i \\
    \textbf{b}^i \\
        \gamma^i \\
    \sigma^i \\
    \end{pmatrix} = \begin{pmatrix}
    \mathcal{I}-\mathcal{K}^{'t} & 0 \\
    \mathcal{J}^{(2)} & \mathcal{P}^{(2)} \\ 
    \mathcal{J}^{(3)} & \mathcal{I}+\mathcal{D}' \\
    \mathcal{J}^{(4)} & -\kappa^2\mathcal{S} \\
    \end{pmatrix}
    \begin{pmatrix}
    \textbf{a} \\
    \gamma \\
    \end{pmatrix}. \\
\end{equation}
\end{subequations}
As is to be expected, the equations are over-determined. But as alluded to in \cite{Li2019DPIE} and discussed in \cite{Eris2022VPIE} and \cite{ref:RothDPIETD2021}, and demonstrated in this paper using analytical basis sets, one can combine potential integral formulations to remove interior resonances. This paper uses combinations that are akin to the EFIE, MFIE, and CFIE. Next, we discuss these equations as well as preconditioners. 

\section{Localized Calder\'{o}n-Type Combined VPIE and SPIE Formulations}

\subsection{Systems of Operators}

From  \eqref{eq:SPIE_pec} it is apparent that one can use either of the two equations but both will have null-spaces at irregular frequencies. The null-spaces of the $\mathcal{S}$ operator correspond to solution of a cavity with Dirichlet boundary conditions whereas those of $\mathcal{I} + \mathcal{D}'$ correspond to those of an interior cavity with Neumann boundary conditions. To avoid these null-spaces, we follow the usual combined field approach, i.e., the two formulations are weighted with a coefficient $\delta$ and $1 - \delta$, where $0 \leq\delta\leq 1$. The Combined SPIE (CSPIE) is written as
\begin{equation}
        \delta\alpha^i + (1-\delta)\beta^i =
        \delta\mathcal{Z}^{SPIE}_{1}\beta + (1 - \delta)\mathcal{Z}^{SPIE}_{2}\beta)
\end{equation}
where
\begin{equation}
\begin{split}
    \mathcal{Z}^{SPIE}_{1} = & ~ \mathcal{S} \\
    \mathcal{Z}^{SPIE}_{2} = & ~ \mathcal{I} + \mathcal{D}'. \\
\end{split}
\end{equation}

Next, we consider the VPIE system. It can be demonstrated rows 1 and 3 of \eqref{eq:VPIE_pec} are spectrally akin to those of an MFIE (denoted by $\mathcal{Z}_1^{VPIE}$) whereas rows 2 and 4 are similar to those of an EFIE (denoted by $\mathcal{Z}_2^{VPIE}$)
\begin{equation}
\begin{split}
    \mathcal{Z}^{VPIE}_{1} = & \begin{pmatrix}
    \mathcal{I} - \mathcal{K}^{'t} & 0 \\
    \mathcal{J}^{(3)} & \mathcal{I} + \mathcal{D}' \\
    \end{pmatrix} \\
    \mathcal{Z}^{VPIE}_{2} = & \begin{pmatrix}
    \mathcal{J}^{(2)} & \mathcal{P}^{(2)} \\
    \mathcal{J}^{(4)} & -\kappa^2\mathcal{S} \\
    \end{pmatrix}.
\end{split}
\end{equation}
The null-spaces of these equations are more complex than those of either the MFIE or EFIE operators, but they share a number of null-spaces. Take $\mathcal{Z}_1^{VPIE}$ for instance. Given the lower triangular nature of the operators, it follows that null-spaces fall into two categories; (a) the null-spaces of  $\mathcal{I} - \mathcal{K}^{'t}$ and (b) null-spaces of $\mathcal{I} + \mathcal{D}'$ for rotational $\mathbf{a}$. While similar analysis is possible for $\mathcal{Z}_2^{VPIE}$ and can be done with the analytical framework given later in the paper, the above is sufficient for the ensuing discussion.

A linear combination of the two systems do not share a null-space. As a result, the Combined VPIE (CVPIE) is prescribed as follows
\begin{equation}
\begin{split}
    \delta
    \begin{pmatrix}
    \textbf{a} \\
    \gamma \\
    \end{pmatrix}^i + (1-\delta)
    \begin{pmatrix}
    \textbf{b} \\
    \sigma \\
    \end{pmatrix}^i = & ~
    \delta~\mathcal{Z}^{VPIE}_{1}
    \begin{pmatrix}
    \textbf{a} \\
    \gamma \\
    \end{pmatrix} \\
    & + (1-\delta)~\mathcal{Z}^{VPIE}_2\begin{pmatrix}
    \textbf{a} \\
    \gamma \\
    \end{pmatrix}. \\ 
\end{split}
\end{equation}
Next, we discuss preconditioners that can be chosen from the identities derived in Section \ref{sec:CalderonIdentities}. 

\subsection{Choosing Analytic Preconditioners\label{sec:precondOps}}

To derive a preconditioner, our approach is to start with the identities presented in \eqref{eq:ScalarCalderonIdentities} and \eqref{eq:VectorCalderonIdentities}. We start with the SPIE. As was alluded to earlier, both $\mathcal{Z}_1^{SPIE}$ and $\mathcal{Z}_2^{SPIE}$ are not unique at irregular frequencies. It is also apparent that  \eqref{eq:ScalarCalderonIdentities} can potentially be used to derive well-conditioned systems. To wit, one can readily recognize that $\mathcal{N}$ operating on $\mathcal{S}$ results in an operator that is second kind, but it is also apparent that such a composition shares null-spaces, i.e.,  $\mathcal{N}\mathcal{S}$ and $\mathcal{I} + \mathcal{D}'$ share resonances due to identity \eqref{eq:NS_identity}. To overcome this conundrum, we choose a complexification of the wavenumber included in $\mathcal{N}$. Specifically, we let $\widetilde{\mathcal{P}}^{SPIE} = -\widetilde{\mathcal{N}}$ such that instead of $\kappa$ we use $\tilde{\kappa} = \kappa - j0.4{H}^{\frac{2}{3}}\kappa^{\alpha}$ \cite{ref:Darbas2006Constant} where $\alpha=\frac{1}{3}$. The resulting Local Calder\'{o}n-Type Combined SPIE (LC-CSPIE) formulation is resonance free and second-kind
\begin{equation} \label{eq: LocalCalderonSPIE}
\begin{split}
        \delta\Tilde{\mathcal{P}}^{SPIE}\alpha^i + (1-\delta)\beta^i = & ~
        \delta\Tilde{\mathcal{P}}^{SPIE}\mathcal{Z}^{SPIE}_{1}[\beta] \\
        & + (1 - \delta)\mathcal{Z}^{SPIE}_{2}[\beta]. 
\end{split}
\end{equation}

Proceeding in a similar manner, we observe that both $\mathcal{Z}^{VPIE}_{1}$ and $\mathcal{Z}^{VPIE}_{2}$ are non-unique at different irregular frequencies. Furthermore, $\mathcal{Z}^{VPIE}_{1}$ is well-conditioned  and $\mathcal{Z}^{VPIE}_{2}$ is ill-conditioned \cite{Li2019DPIE}. As before, \eqref{eq:VectorCalderonIdentities} provides the necessary fodder for developing a preconditioner. Note, that $\mathcal{Z}^{VPIE}_{1}$ is well-conditioned \cite{Li2019DPIE}, wheres $\mathcal{Z}^{VPIE}_{2}$ is ill-conditioned \cite{Li2019DPIE}. Using \eqref{eq:VectorCalderonIdentities} consider the following preconditioning operator and its action on $\mathcal{Z}^{VPIE}_{2}$
\begin{equation} \label{eq:CalderonPrecond}
    \begin{pmatrix}
    \kappa^2\mathcal{L}^{t} & -\mathcal{Q}^{(1)} \\
    \mathcal{M}^{(3)}  & -\mathcal{Q}^{(3)} \\
    \end{pmatrix}
    \begin{pmatrix}
    \mathcal{J}^{(2)} & \mathcal{P}^{(2)} \\
    \mathcal{J}^{(4)} & -\kappa^2\mathcal{S} \\
    \end{pmatrix}.
\end{equation}
It can be shown using \eqref{NewCalderon_a}, \eqref{NewCalderon_c}, \eqref{NewCalderon_i}, and \eqref{NewCalderon_k} that \eqref{eq:CalderonPrecond} reduces to
\begin{equation} \label{eq: PrecondVPIE}
    \begin{pmatrix}
    -(\mathcal{I}-\mathcal{K}^{'t})\mathcal{K}^{'t} & 0 \\
    -\mathcal{J}^{(3)}\mathcal{K}^{'t} + (\mathcal{I} + \mathcal{D}')\mathcal{J}^{(3)} & (\mathcal{I} + 
    \mathcal{D}')\mathcal{D}'\\
    \end{pmatrix}.
\end{equation}
It is apparent that \eqref{eq: PrecondVPIE} includes an identity plus compact operator along the diagonal. The off-diagonal operators are compact operators acting on compact operators or compact operators acting on bounded operators, the result of both being compact \cite{Rudin1991FA}. As a result, the entire system of operators can be partitioned into identity plus compact operators, which are well-conditioned by definition. While this is a desirable outcome, \eqref{eq: PrecondVPIE} has null-spaces that pose challenges.  

Examining \eqref{eq: PrecondVPIE} in manner similar to what was done for $\mathcal{Z}_1^{VPIE}$, the lower triangular nature of the operators dictate that the null-spaces fall into two categories; (a) those of $(\mathcal{I}-\mathcal{K}^{'t})\mathcal{K}^{'t}$ and (b) for any rotational $\mathbf{a}$ the null-spaces of $(\mathcal{I} + \mathcal{D}')\mathcal{D}'$. It is apparent that these null-spaces are shared with $\mathcal{Z}_1^{VPIE}$. Furthermore, the below factorization demonstrates $\mathcal{Z}_1^{VPIE}$ and \eqref{eq: PrecondVPIE} share nullspaces,


\begin{equation} \label{eq: Factor_PrecondVPIE}
    \begin{split}
        \begin{pmatrix}
            -(\mathcal{I}-\mathcal{K}^{'t}) & 0 \\
            -\mathcal{J}^{(3)}  & \mathcal{I} + 
            \mathcal{D}'\\
        \end{pmatrix}\begin{pmatrix}
            \mathcal{K}^{'t} & 0 \\
            \mathcal{J}^{(3)} & \mathcal{D}'\\
        \end{pmatrix}.
    \end{split}
\end{equation}
An alignment of these operators' nullspaces is numerically depicted in Fig. 4.


As before, to ensure uniqueness at all frequencies, the preconditioned operators are complexified. Specifically, $\tilde{\kappa}$ from \cite{ref:Darbas2006Constant} is again selected to localize the preconditioner and interlace the nullspaces of the preconditioned $\mathcal{Z}^{VPIE}_{2}$ and  $\mathcal{Z}^{VPIE}_{1}$ operators. The new Local Calder\'{o}n-Type Combined VPIE (LC-CVPIE) formulation is
\begin{equation} \label{eq: LocalCalderonVPIE}
\begin{split}
    \delta
    \begin{pmatrix}
    \textbf{a} \\
    \gamma \\
    \end{pmatrix}^i + & ~(1-\delta)\Tilde{\mathcal{P}}^{VPIE}
    \begin{pmatrix}
    \textbf{b} \\
    \sigma \\
    \end{pmatrix}^i = ~
    \delta~\mathcal{Z}^{VPIE}_{1}
    \begin{pmatrix}
    \textbf{a} \\
    \gamma \\
    \end{pmatrix} \\
    & + (1-\delta)~\Tilde{\mathcal{P}}^{VPIE}\mathcal{Z}^{VPIE}_2\begin{pmatrix}
    \textbf{a} \\
    \gamma \\
    \end{pmatrix} \\ 
\end{split}
\end{equation}
where
\begin{equation}
    \Tilde{\mathcal{P}}^{VPIE} = -\begin{pmatrix}     \tilde{\kappa}^2\tilde{\mathcal{L}}^{t} & -\tilde{\mathcal{Q}}^{(1)} \\
    \tilde{\mathcal{M}}^{(3)}  & -\tilde{\mathcal{Q}}^{(3)} \\
    \end{pmatrix}.
\end{equation}

Next, we discuss the implementation of these equation using both an analytical framework as well as basis functions defined on piecewise tesselation. Of course, the LC-CVPIE and LC-CSPIE together constitute the Local Calder\'{o}n Combined Decoupled Potential Integral Equation (LC-CDPIE) whose unknowns are the concatenation of the unknowns for the VPIE and SPIE constrained to PEC objects. 

\subsection{Left and Right Preconditioners}

Left and right preconditioners are used to scale the various operators in the suggested VPIE formulation such that all unknowns are of the same units. The left and right scaling matrices are detailed in \cite{Li2019DPIE}\cite{ref:Baumann2022DPIE}. 

For the LC-CVPIE in \eqref{eq: LocalCalderonVPIE}, these matrices are
\begin{subequations} \label{eq: VPIE_scalings}
    \begin{align}
        \mathcal{P}_l^{(1)} = ~ & \text{diag}\left(1, -j\kappa\right) \\
        \mathcal{P}_l^{(2)} = ~ & \text{diag}\left(-j\kappa, ~1\right) \\
        \mathcal{P}_r^{(1)} = ~ & \left(\mathcal{P}_l^{(1)}\right)^{-1} \\
        \mathcal{P}_r^{(2)} = ~ & \left(\mathcal{P}_l^{(2)}\right)^{-1}.
    \end{align}
\end{subequations}

 After scaling \eqref{eq: LocalCalderonVPIE} with \eqref{eq: VPIE_scalings}, the new well-conditioned formulation implemented in the results section is
\begin{equation}
\begin{split}
    & \delta
    \mathcal{P}_l^{(1)}\begin{pmatrix}
    \textbf{a} \\
    \gamma \\
    \end{pmatrix}^i + (1-\delta)\mathcal{P}_l^{(1)}\Tilde{\mathcal{P}}^{VPIE}\mathcal{P}_r^{(2)}\mathcal{P}_l^{(2)}
    \begin{pmatrix}
    \textbf{b} \\
    \sigma \\
    \end{pmatrix}^i = \\
    & ~~~\delta\mathcal{P}_l^{(1)}\mathcal{Z}^{VPIE}_{1}\mathcal{P}_r^{(1)}
    \begin{pmatrix}
    \Tilde{\textbf{a}} \\
    \Tilde{\gamma} \\
    \end{pmatrix} \\
    & ~~~ ~~~ + (1-\delta)~\mathcal{P}_l^{(1)}\Tilde{\mathcal{P}}^{VPIE}\mathcal{P}_r^{(2)}\mathcal{P}_l^{(2)}\mathcal{Z}^{VPIE}_{2}\mathcal{P}_r^{(1)}
    \begin{pmatrix}
    \Tilde{\textbf{a}} \\
    \Tilde{\gamma} \\
    \end{pmatrix} \\ 
\end{split}
\end{equation}
where
\begin{equation}
    \begin{pmatrix}
        \Tilde{\textbf{a}} \\
        \Tilde{\gamma} \\
    \end{pmatrix} = \mathcal{P}_l^{(1)}\begin{pmatrix}
        \textbf{a} \\
        \gamma \\      
    \end{pmatrix}. \\
\end{equation}

For the LC-CSPIE, the scaling parameters are
\begin{subequations}
\begin{align}
    \mathcal{S}^{(1)}_l = & -j\kappa \\
    \mathcal{S}^{(2)}_l = & 1
\end{align}
\end{subequations}
which results in the following LC-CSPIE formulation,

\begin{equation}
\begin{split}
    &\delta \mathcal{S}_l^{(2)}\mathcal{P}^{SPIE}\mathcal{S}_r^{(1)}\mathcal{S}_l^{(1)}\alpha^i + (1-\delta)\mathcal{S}^{(2)}_l\beta^i = \\
    & ~~~ \delta \mathcal{S}_l^{(2)}\mathcal{P}^{SPIE}\mathcal{S}_r^{(1)}\mathcal{S}_l^{(1)}\mathcal{Z}^{SPIE}_{1}\mathcal{S}_r^{(2)}[\Tilde{\beta}] \\ 
    & ~~~ + (1 - \delta)\mathcal{S}_l^{(2)}\mathcal{Z}^{SPIE}_{2}\mathcal{S}_r^{(2)}[\Tilde{\beta}]
\end{split}
\end{equation}
where

\begin{equation}
    \Tilde{\beta} = \mathcal{S}^{(2)}_l\beta.
\end{equation}

\subsection{Mapping Properties}

The DPIE operators in \eqref{eq:VPIEsystem} and \eqref{eq:SPIEsystem} are a map between function spaces. We now summarize these mapping properties using vector and scalar spherical harmonic functions defined in Appendix B for a collection of domains and associate ranges 

\begin{subequations}
    \begin{align}
    \mathcal{Z}^{VPIE}:& ~\begin{pmatrix}
    \mathbf{\Psi} \\
    0 \\
    0 \\
    \end{pmatrix}\longrightarrow\begin{pmatrix}
    \mathbf{\Psi}~\text{or}~ 0\\
    0 \\
    0 \\
    \end{pmatrix} \\
    \mathcal{Z}^{VPIE}:& ~\begin{pmatrix}
    0 \\
    \mathbf{\Phi} \\
    0 \\
    \end{pmatrix}\longrightarrow\begin{pmatrix}
    0 \\
    \mathbf{\Phi}~\text{or}~ Y \\
    0 \\
    \end{pmatrix} \\
    \mathcal{Z}^{VPIE}:& ~\begin{pmatrix}
    0 \\
    0 \\
    Y \\
    \end{pmatrix}\longrightarrow\begin{pmatrix}
    0 \\
    0 \\
    \mathbf{\Phi} \\
    \end{pmatrix} \\
    \mathcal{Z}^{SPIE}:& ~~Y\longrightarrow Y.
    \end{align}
\end{subequations}

These mapping properties demonstrate that div-conforming and curl-conforming spaces map to div-conforming and curl-conforming spaces, respectively. Therefore, the Gram matrix of a MOM system with RWG and hat testing and basis functions is well-conditioned for the DPIE operators and non-degenerate. 

\section{Discrete Implementation}

The crux of what follows is a means of discretizing \eqref{eq: LocalCalderonSPIE} and \eqref{eq: LocalCalderonVPIE}. It is apparent that to do so one needs both scalar and vector basis. For analytical analysis, the basis are spherical and vector spherical harmonics. For piecewise tesselation, these are the $W^0$ or hat functions and $W^2$ or Rao-Wilton-Glisson (RWG) functions. It is important to keep in mind that for \eqref{eq: LocalCalderonVPIE} we are \emph{not} representing the vector potentials; instead, we are representing either traces of or operators acting on the vector potential. Likewise, for \eqref{eq: LocalCalderonSPIE}, we represent the potential and its normal derivative on the manifold.  

\subsection{Analytic Basis Sets}

Using the notation in \cite{ref:Baumann2022DPIE}, the spherical harmonic basis functions are denoted by $\underline{\mathcal{B}}^s_n$ where $n\in [0,N_s)$ and the vector harmonic basis functions are denoted by $\underline{\mathcal{B}}^v_n$ where $n\in [0,N_v)$. The number of basis functions in \eqref{eq: LocalCalderonSPIE} is $N_s$, and the span is denoted by $\overline{\tau}^{\text{SPIE}} = \sum_{n=0}^{N_s - 1} \underline{\mathcal{B}}^s_n y^{\text{SPIE}}_n$ where $y^s_n$ are basis function coefficients. The number of basis functions in \eqref{eq: LocalCalderonVPIE} is $N_s + N_v$.  Likewise, the span of basis functions is denoted by $\overline{\tau}^{\text{VPIE}} = \sum_{n=0}^{N_s + N_v - 1} \underline{\mathcal{F}}^V_n y^{\text{VPIE}}_n$ where $y^v_n$ is a list of basis function coefficients and $\underline{\mathcal{F}}^V = \text{diag}\left( \underline{\mathcal{B}}^s_n, \underline{\mathcal{B}}^v_n \right)$. 

The sets of harmonics are denoted by
\begin{subequations}
\begin{align}
    \underline{\mathcal{B}}_Y = & \begin{pmatrix}
    Y^0_0 & \dots & Y^m_l & \dots & Y^{N_h}_{N_h}
    \end{pmatrix} \\
    \underline{\mathcal{B}}_{\mathbf{\Psi}} = & \begin{pmatrix}
    \mathbf{\Psi}^0_0 & \dots & \mathbf{\Psi}^m_l & \dots & \mathbf{\Psi}^{N_h}_{N_h}
    \end{pmatrix} \\
    \underline{\mathcal{B}}_{\mathbf{\Phi}} = & \begin{pmatrix}
    \mathbf{\Phi}^0_0 & \dots & \mathbf{\Phi}^m_l & \dots & \mathbf{\Phi}^{N_h}_{N_h}
    \end{pmatrix}
\end{align}
\end{subequations}
and the scalar and vector basis functions are then
\begin{subequations}
\begin{align}
    \underline{\mathcal{B}}^s = & ~ \underline{\mathcal{B}}_Y \\
    \underline{\mathcal{B}}^v = & \begin{pmatrix}
    \underline{\mathcal{B}}_{\mathbf{\Phi}} & \underline{\mathcal{B}}_{\mathbf{\Psi}}
    \end{pmatrix}
\end{align}
\end{subequations}
where $N_s = \left(N_h + 1 \right)^2$ and $N_v = 2 \left(N_h + 1 \right)^2$ and $N_h = \kappa a + 2$ where $a$ is the radius of the sphere. 

\subsection{Piece-Wise Basis Sets}

The SPIE and VPIE discrete scalar and vector basis functions are hat and RWG functions on a mesh with $N_f$ faces and $N_e$ edges and $N_n$ nodes. The hat functions are defined by
\begin{subequations}
\begin{align}
    h_{n_n} \left( \textbf{r} \right)  = & 
    \begin{cases}
    \frac{l_{n_n}}{2 A_{n_n}}\hat{\mathbf{u}}_{n_n}\cdot\left(\textbf{e}_{n_n} - \boldsymbol{\rho}^+_{n_n} \right) & \textbf{r}\in T_{n_n} \\
    0 & \text{else} \\
    \end{cases}    
\end{align}
\end{subequations}
and the RWG functions are defined by
\begin{subequations}
\begin{align}
    \mathbf{f}_{n_e} \left( \textbf{r} \right)  = & 
    \begin{cases}
    \frac{l_{n_e}}{2 A_{n_e}^{\pm}}\boldsymbol{\rho}^{\pm}_{n_e} & \textbf{r}\in T^{\pm}_{n_e} \\
    0 & \text{else} \\
    \end{cases}    
\end{align}
\end{subequations}
where
\begin{subequations}
\begin{align}
    \boldsymbol{\rho}^{\pm} \left( \textbf{r} \right)  = & 
    \begin{cases}
    \pm(\textbf{r} - \textbf{p}^{\pm}_n) & \textbf{r}\in T^{\pm}_{n} \\
    0 & \text{else} \\
    \end{cases}    
\end{align}
\end{subequations}
and $A_{n_n}$ is the area of a face connected to node $n_n$; $A_{n_e}^{\pm}$ is the area of the faces sharing edge $n_e$; $\hat{\mathbf{u}}_{n_n}$ is the  planar normal of the edge opposite of node $n_n$ on face $\text{T}_{n_n}$ and pointing away from node $n_n$; $l_{n_n}$ is the length of the edge opposite of node $n_n$ and on face $\text{T}_{n_n}$; $l_{n_e}$ is the length of the edge $n_e$; $\boldsymbol{\rho}^{\pm}_{n_n}$ originates in node $n_n$ and terminates in face $\text{T}_{n_n}$; $\textbf{e}_{n_n}$ are the coordinates of either node on the edge opposite of node $n_n$; $\text{T}^{\pm}_{n_e}$ are the faces sharing edge $n_e$; $\text{T}^{\pm}_{n}$ are either nodal or edge faces; $\boldsymbol{\rho}^{\pm}_{n_e}$ are the coordinates for the nodes opposite of the selected edge on the appropriate face sharing the selected edge; and $\textbf{p}^{\pm}_{n}$ are the appropriate nodal coordinates for the RWG or hat basis. The scalar and vector basis functions are then
\begin{subequations}
\begin{align}
    \underline{\mathcal{B}}^s = & \begin{pmatrix}
    h_{0} & \dots & h_{n_n} & \dots & h_{N_n}
    \end{pmatrix} \\
    \underline{\mathcal{B}}^v = & \begin{pmatrix}
    \mathbf{f}_{0} & \dots & \mathbf{f}_{n_e} & \dots & \mathbf{f}_{N_e}
    \end{pmatrix}.
\end{align}
\end{subequations}


\subsection{Discrete System}

The Method of Moments (MOM) system is constructed with Galerkin testing using inner products defined as
\begin{subequations}
	\begin{align}
		\left\langle g \left( \mathbf{r} \right), f \left( \mathbf{r} \right) \right\rangle =& \int g^* \left( \mathbf{r} \right) f \left( \mathbf{r} \right) dS \\
		\left\langle \mathbf{g} \left( \mathbf{r} \right), \mathbf{f} \left( \mathbf{r} \right) \right\rangle =& \int \mathbf{g}^* \left( \mathbf{r} \right) \cdot \mathbf{f} \left( \mathbf{r} \right) dS.
	\end{align}
\end{subequations}

The LC-CVPIE discrete system is

\begin{equation}
    \mathbb{Z}^{\text{VPIE}}\Tilde{y}^{\text{VPIE}} = b^{\text{VPIE}}
\end{equation}
where the elements are defined by
\begin{subequations}
    \begin{align}
        & \mathbb{Z}_{kn}^{\text{VPIE}} = ~ \langle \underline{\mathcal{F}}^V_k, ~ \delta\text{G}^{-1}\mathcal{P}_l^{(1)}\mathcal{Z}^{VPIE}_{1}\mathcal{P}_r^{(1)}\underline{\mathcal{F}}^V_n \\
         & ~~~~ + (1-\delta)~\text{G}^{-1}\mathcal{P}_l^{(1)}\Tilde{\mathcal{P}}^{VPIE}\mathcal{P}_r^{(2)}\text{G}^{-1}\mathcal{P}_l^{(2)}\mathcal{Z}^{VPIE}_2\mathcal{P}_r^{(1)}\underline{\mathcal{F}}^V_n \rangle \\
        & \tilde{y}^{\text{VPIE}}_n = ~ \mathcal{P}_l^{(1)} y_n^{\text{VPIE}} \\
        & b_{k}^{\text{VPIE}} = ~ \langle \underline{\mathcal{F}}^V_k,\delta
        \text{G}^{-1}\mathcal{P}_l^{(1)}\begin{pmatrix}
        \textbf{a} \\
        \gamma \\
        \end{pmatrix}^i \\
        & ~~~ + (1-\delta)\text{G}^{-1}\mathcal{P}_l^{(1)}\Tilde{\mathcal{P}}^{VPIE}\mathcal{P}_r^{(2)}\text{G}^{-1}\mathcal{P}_l^{(2)}
        \begin{pmatrix}
        \textbf{b} \\
        \sigma \\
        \end{pmatrix}^i \rangle \\
        & \text{G}_{kn} = ~ \langle \underline{\mathcal{F}}^V_k, \underline{\mathcal{F}}^V_n \rangle.
    \end{align}
\end{subequations}

The LC-CSPIE discrete system is

\begin{equation}
    \mathbb{Z}^{\text{SPIE}}y^{\text{SPIE}} = \text{b}^{\text{SPIE}}
\end{equation}
where the elements are defined by
\begin{subequations}
    \begin{align}
        \mathbb{Z}_{kn}^{\text{SPIE}} = & ~ \langle \underline{\mathcal{B}}^s_k, ~ \delta G^{-1}\mathcal{S}_l^{(2)}\mathcal{P}^{SPIE}\mathcal{S}_r^{(1)}G^{-1}\mathcal{S}_l^{(1)}\mathcal{Z}^{SPIE}_{1}\mathcal{S}_r^{(2)}\underline{\mathcal{B}}^s_n \\
         & ~~~ + (1 - \delta)G^{-1}\mathcal{S}_l^{(2)}\mathcal{Z}^{SPIE}_{2}\mathcal{S}_r^{(2)}\underline{\mathcal{B}}^s_n \rangle \\
        b_{k}^{\text{SPIE}} = & ~ \langle \underline{\mathcal{B}}^s_k,\delta G^{-1}\mathcal{S}_l^{(2)}\mathcal{P}^{SPIE}\mathcal{S}_r^{(1)}G^{-1}\mathcal{S}_l^{(1)}\alpha^i \\
        & ~~~ + (1-\delta)\text{G}^{-1}\mathcal{S}^{(2)}_l\beta^i \rangle \\
        \text{G}_{kn} = & ~ \langle \underline{\mathcal{B}}^s_k, \underline{\mathcal{B}}^s_n \rangle.
    \end{align}
\end{subequations}

A few notes are in order. First, the above integrations use singularity subtraction to handle the singularity region where the testing function is within $0.15\lambda$ of the basis function as detailed in \cite{ref:Baumann2022DPIE}.

\subsection{Zero-Mean Constraint}

Both $\gamma = \hat{\textbf{n}}\cdot\textbf{A}(\textbf{r})$ and $\beta = \hat{\textbf{n}}\cdot\nabla\phi(\textbf{r})$ have a zero-mean constraint (ZMC) \cite{Li2019DPIE}. This is enforced with two different methods. Method 1 is a rank-one update technique embedding the Lagrange multiplier in the forward Galerkin tested system by appending a list of basis function areas (entries corresponding to non-zero-mean constrained variables are zero) for $\gamma$ and $\beta$ unknowns to the rows and columns of the forward matrix \cite{ref:Baumann2022DPIE}. Method 2 solves the same VPIE and SPIE Galerkin tested system with a RHS of incident potentials and a RHS of basis function areas (entries corresponding to non-zero-mean constrained variables are zero) as discussed in \cite{ref:Dault2015Penalty}.

\subsection{Fine-Grain Localization}\label{Sec:Localization}

Localizing \eqref{eq: LocalCalderonSPIE} and \eqref{eq: LocalCalderonVPIE} with a single complex wavenumber determined by the geometry's global max-mean curvature is not necessary and, in some cases, sub-optimal. A simple generalization of this localization approach is computing the max-mean curvature of each edge and defining the max- mean curvature of a patch as the maximum of the mean curvatures of the patch's edges. Note, when computing the discrete preconditioner, the RWG and hat interactions are constituted by the interactions of the patches composing the basis and testing functions. Then, for each source-observer patch pair, the localizing complex wavenumber is generated from the maximum of the max-mean curvatures of the sources and observers. Here, this is referred to as fine-grain localization, and we use this approach unless otherwise stated.

\section{Results}

In what follows, we will demonstrate that (a) the new formulations \eqref{eq: LocalCalderonSPIE} and \eqref{eq: LocalCalderonVPIE} have flat, low eigenvalue conditioning from low-frequency to high-frequency, (b) the suggested SPIE in \eqref{eq: LocalCalderonSPIE} also has flat, low singular value conditioning across the band while \eqref{eq: LocalCalderonVPIE} increases at low-frequency and (c) the preconditioned VPIE and SPIE formulations rapidly converge to potential quantities which generate the correct RCS plots for spherical,  multi-scale, and arbitrary geometries in comparison to the standard CFIE. In all examples that follow, $\delta$ is chosen to be 0.5. This is very similar to the choice of weight for constructing the CFIE.

\subsection{Spectral Properties with Analytic Basis Sets}

\begin{figure}[h]
\includegraphics[width=8cm]{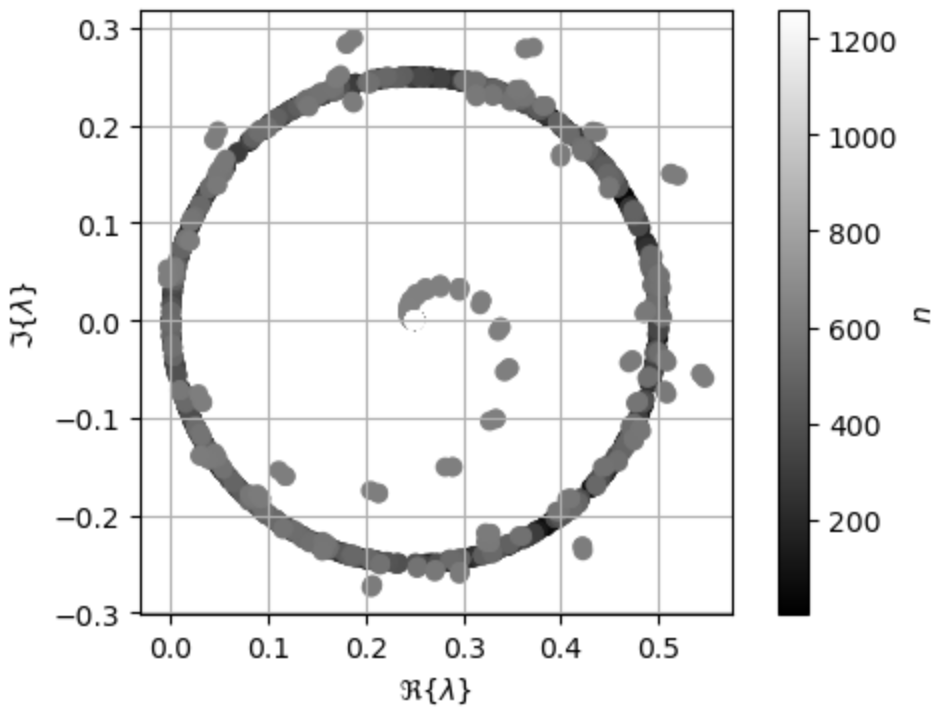}
\centering
\centering\caption{The spectrum of (\ref{eq: forward1}). Eigenvalues are denoted by \text{\emph{n}}. }
\label{fig:EigCalderonVPIE}
\end{figure}

\begin{figure}[h]
\includegraphics[width=8cm]{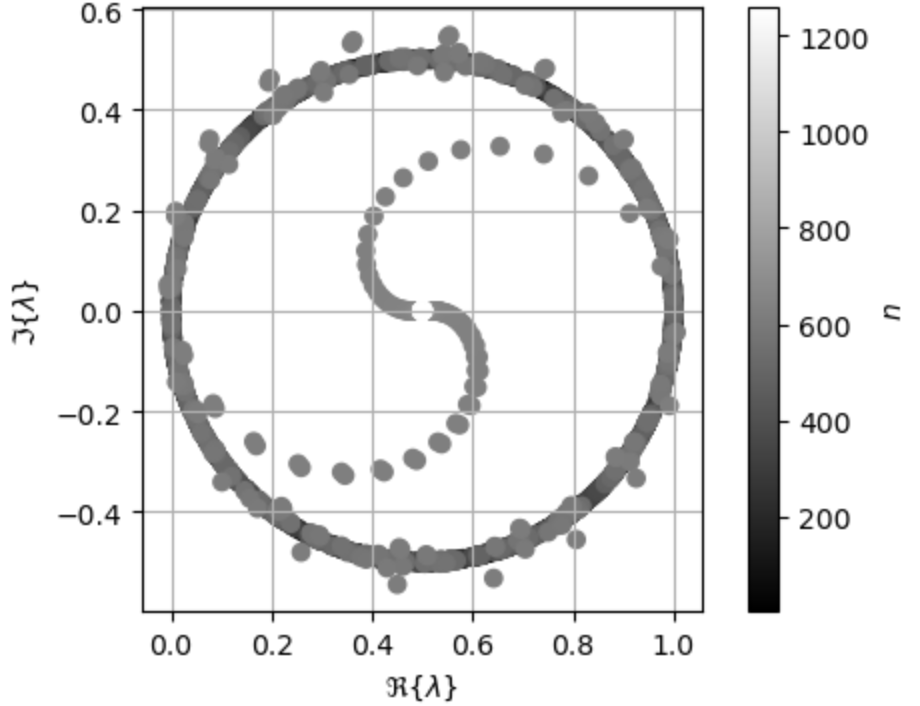}
\centering
\centering\caption{The spectrum of (\ref{eq: forward2}). Eigenvalues are denoted by \text{\emph{n}}.}
\label{fig:Eig13VPIE}
\end{figure}

\begin{figure}[h]
\includegraphics[width=8cm]{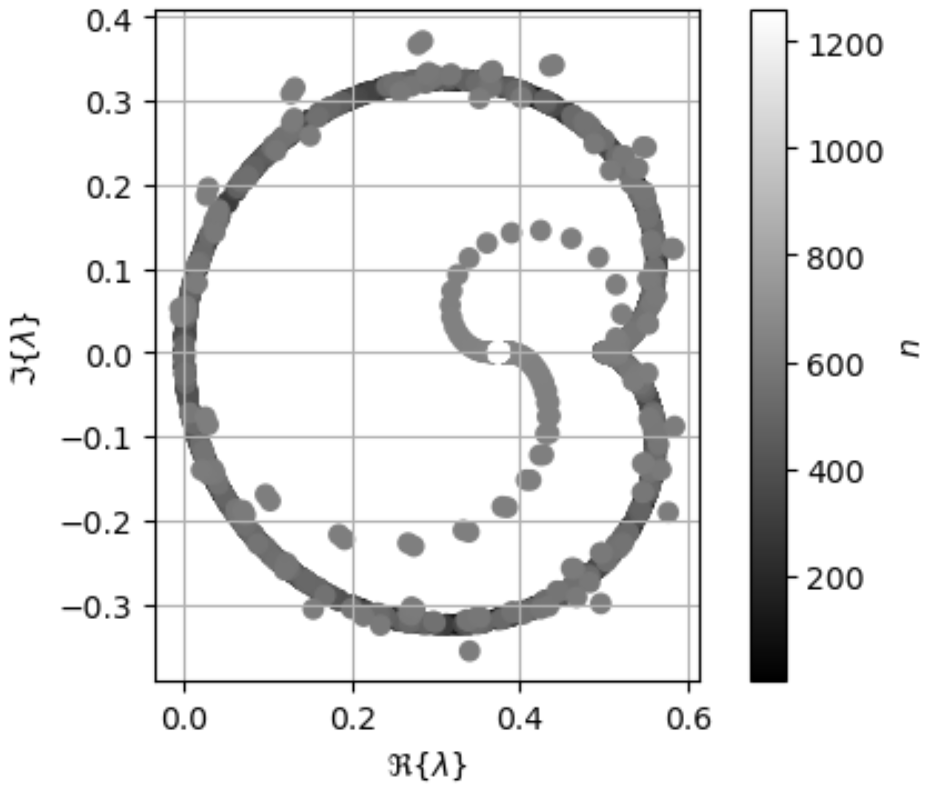}
\centering
\centering\caption{The spectrum of (\ref{eq: forward3}). Eigenvalues are denoted by \text{\emph{n}}.}
\label{fig:EigCombinedCalderonVPIE}
\end{figure}

We now analyze the generalized eigenvalues of the MOM system for the various, suggested formulations. The eigenvalue condition number is denoted by $\kappa(\lambda, \mathcal{Z})$ and defined as the ratio of maximum and minimum eigenvalue moduli. The singular value condition number is denoted by $\kappa(\mathcal{Z})$. Figures \ref{fig:EigCalderonVPIE}-\ref{fig:EigCombinedCalderonVPIE} show the eigenvalues of
\begin{subequations}
    \begin{align}
        & G^{-1}\mathcal{P}_l^{(1)}\mathcal{P}^{VPIE}\mathcal{P}_r^{(2)} G^{-1} \mathcal{P}_l^{(2)}\mathcal{Z}^{VPIE}_2\mathcal{P}_r^{(1)}\label{eq: forward1} \\
        & G^{-1}\mathcal{P}_l^{(1)}\mathcal{Z}^{VPIE}_1\mathcal{P}_r^{(1)}\label{eq: forward2} \\
        &\delta G^{-1}\mathcal{P}_l^{(1)}\mathcal{Z}^{VPIE}_1\mathcal{P}_r^{(1)} \notag \\ & ~~~ + (1-\delta)G^{-1}\mathcal{P}_l^{(1)}\mathcal{P}^{VPIE}\mathcal{P}_r^{(2)} G^{-1} \mathcal{P}_l^{(2)}\mathcal{Z}^{VPIE}_2\mathcal{P}_r^{(1)}\label{eq: forward3}
    \end{align}
\end{subequations}
and respectively for a unit sphere illuminated at 30 GHz. Each operator has bounded eigenvalues but some eigenvalues are near the origin of the complex plane. 

\begin{figure}[h]
\includegraphics[width=8cm]{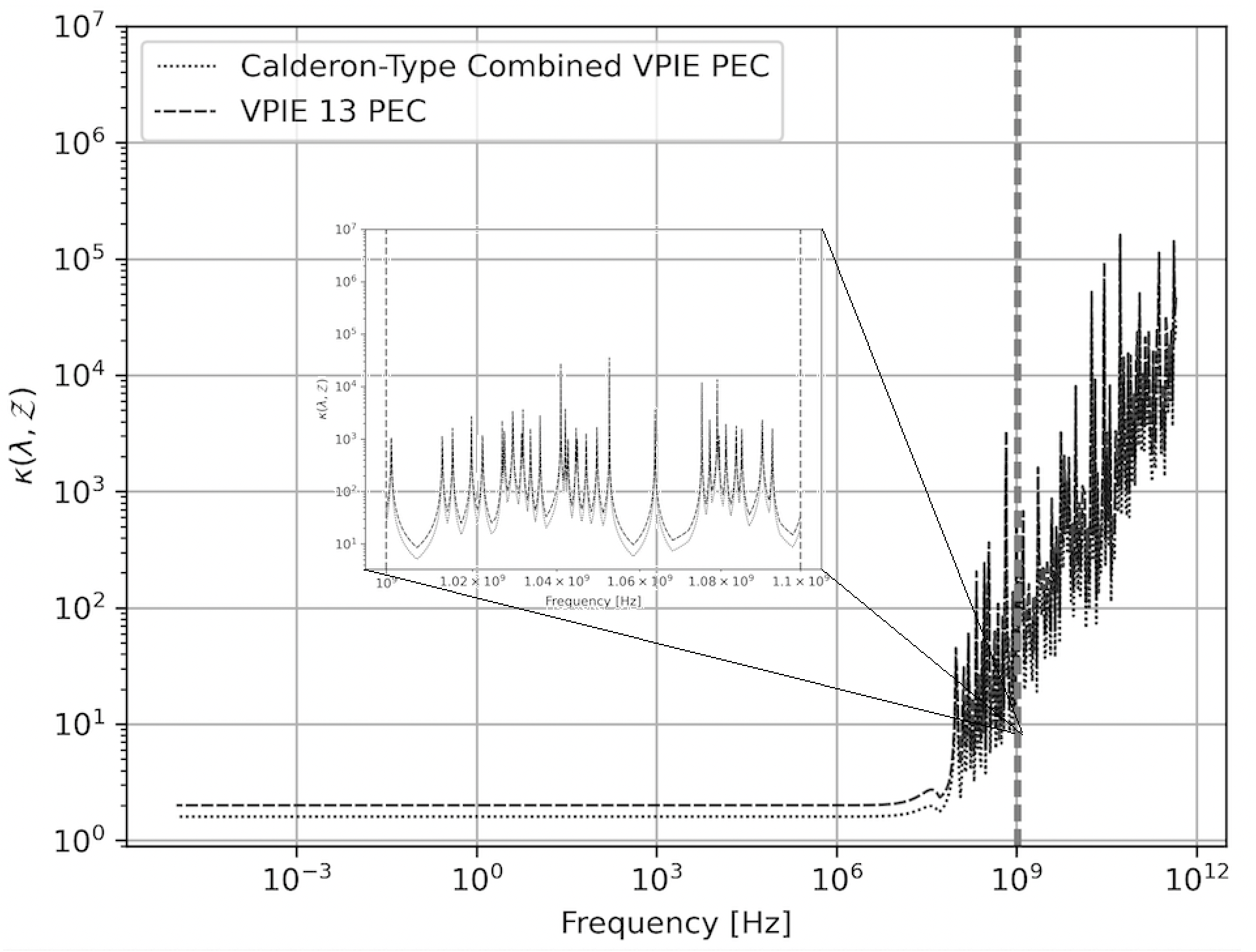}
\centering
\centering\caption{Plot of eigenvalue conditioning.}
\label{fig:CondCombinedCalderonVPIE}
\end{figure}

Figure \ref{fig:CondCombinedCalderonVPIE} depicts the conditioning for frequencies between 1e-5 Hz and 1e12 Hz for formulations (\ref{eq: forward2}) and (\ref{eq: forward3}). The sharp increases in condition number aligns at all shown frequencies. This alignment indicates the two formulations share the same resonances, and, therefore, Fig. \ref{fig:CondCombinedCalderonVPIE} numerically verifies the properties of new Calder\'{o}n-type identities developed in Section \ref{sec:precondOps}. Also, the condition numbers of formulations (\ref{eq: forward2}) and (\ref{eq: forward3}) increases $\propto\omega$ in the high-frequency region. The more common Calder\'{o}n preconditioning approach for the CFIE and the dielectric implementation of the DPIE show similar results \cite{ref:Cools2009Calderon}\cite{Hsiao97calderon}\cite{ref:Baumann2022DPIE}.


\begin{figure}[h]
\includegraphics[width=8.5cm]{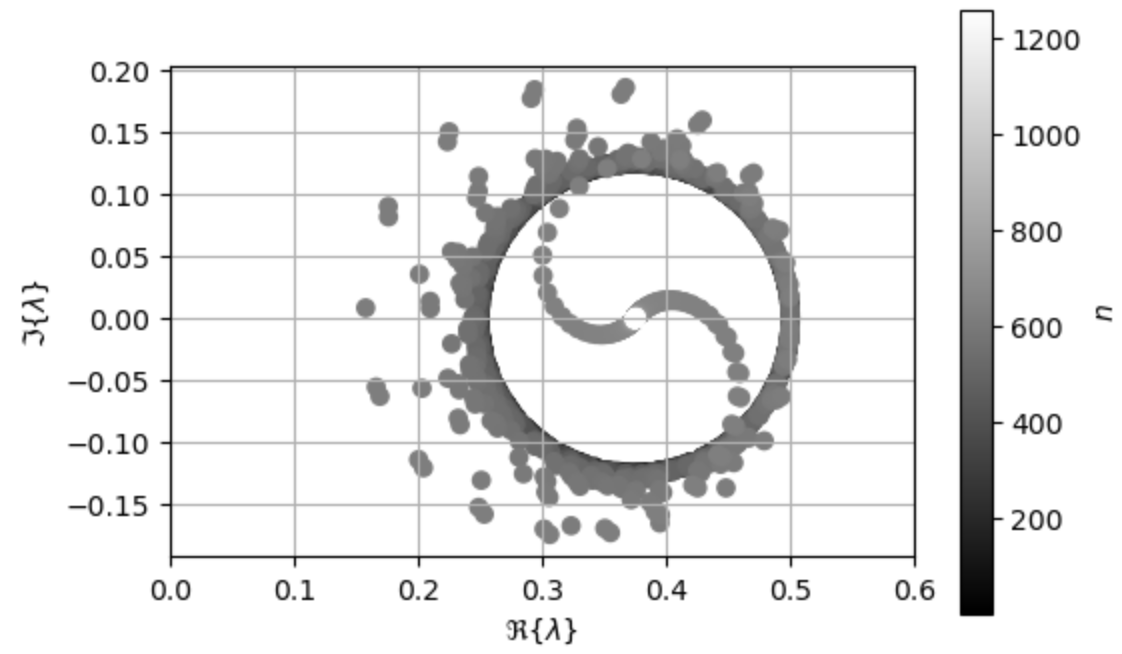}
\centering
\centering\caption{Spectrum of Local Calder\'{o}n-Type Combined VPIE formulation. Eigenvalues are denoted by \text{\emph{n}}.}
\label{fig:EigLocalCalderonCombinedVPIE}
\end{figure}

\begin{figure}[h]
\includegraphics[width=8cm]{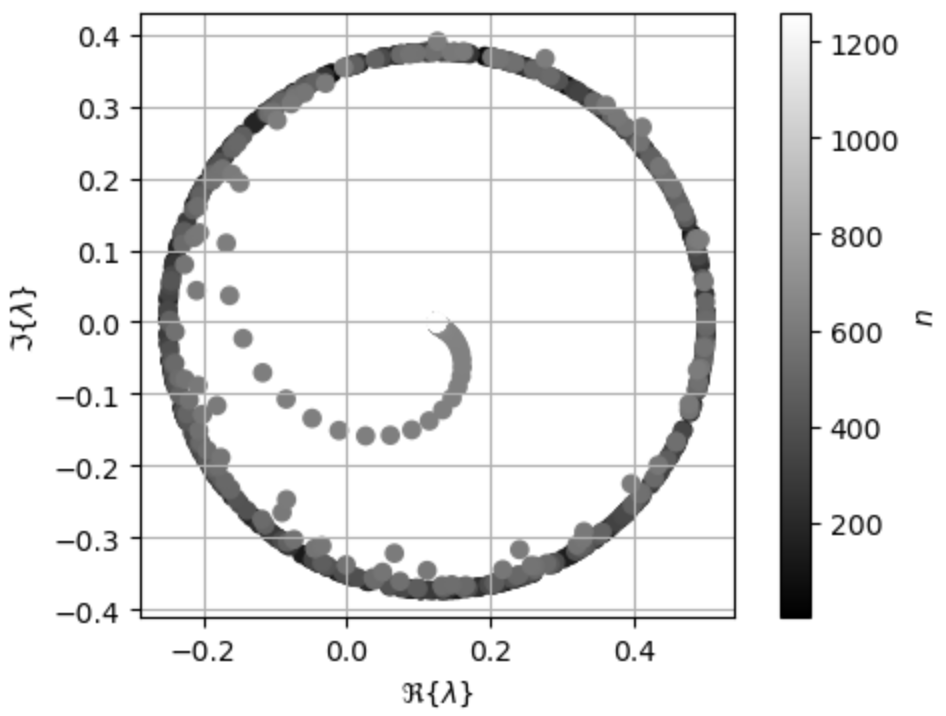}
\centering
\centering\caption{Spectrum of Local Calder\'{o}n-Type Combined SPIE formulation. Eigenvalues are denoted by \text{\emph{n}}.}
\label{fig:EigLocalCalderonCombinedSPIE}
\end{figure}

Next, Fig. \ref{fig:EigLocalCalderonCombinedVPIE} and Fig. \ref{fig:EigLocalCalderonCombinedSPIE} show the eigenvalues for the LC-CVPIE and LC-CSPIE formulations, respectively, for a unit sphere illuminated at 30 GHz. The eigenvalues are bounded and collected away from the origin unlike the spectra in Fig. \ref{fig:EigCalderonVPIE}-\ref{fig:EigCombinedCalderonVPIE}. The LC-CSPIE has an eigenvalue condition number of 4 and singular value condition number of 6. The spectrum of the LC-CVPIE has an eigenvalue condition number of 4 and singular value condition number of 4. 

\begin{figure}[h]
\includegraphics[width=8cm]{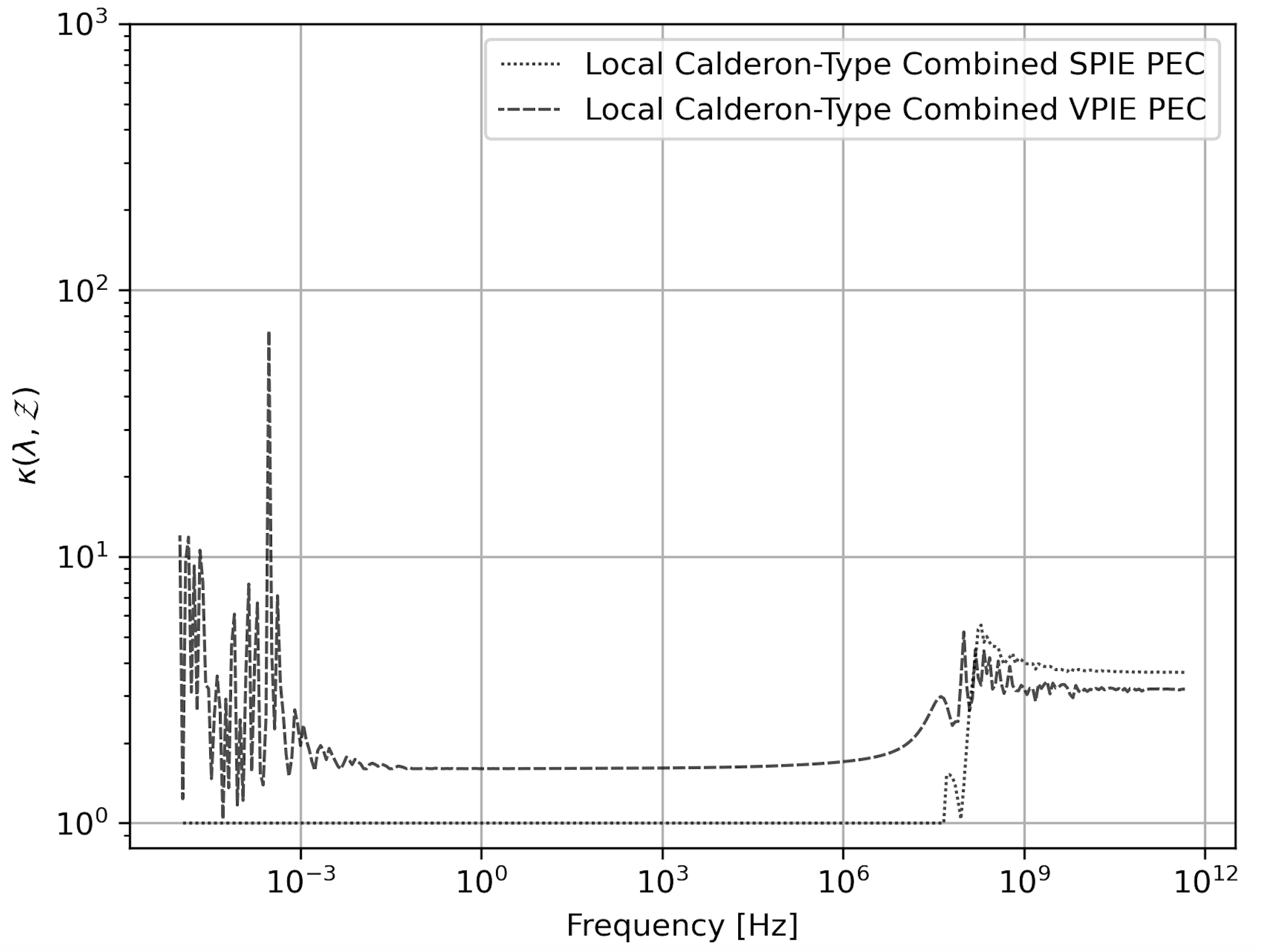}
\centering
\centering\caption{Plot of eigenvalue conditioning.}
\label{fig:EigValCond}
\end{figure}

Figure \ref{fig:EigValCond} is a eigenvalue condition number plot for the LC-CVPIE and LC-CSPIE formulations  for frequencies between 1e-5 Hz and 1e12 Hz. The condition number is flat at low-frequency and does not increase $\propto\omega$ in the high-frequency region unlike the formulations in Fig. \ref{fig:CondCombinedCalderonVPIE}.
\begin{figure}[h]
\includegraphics[width=8cm]{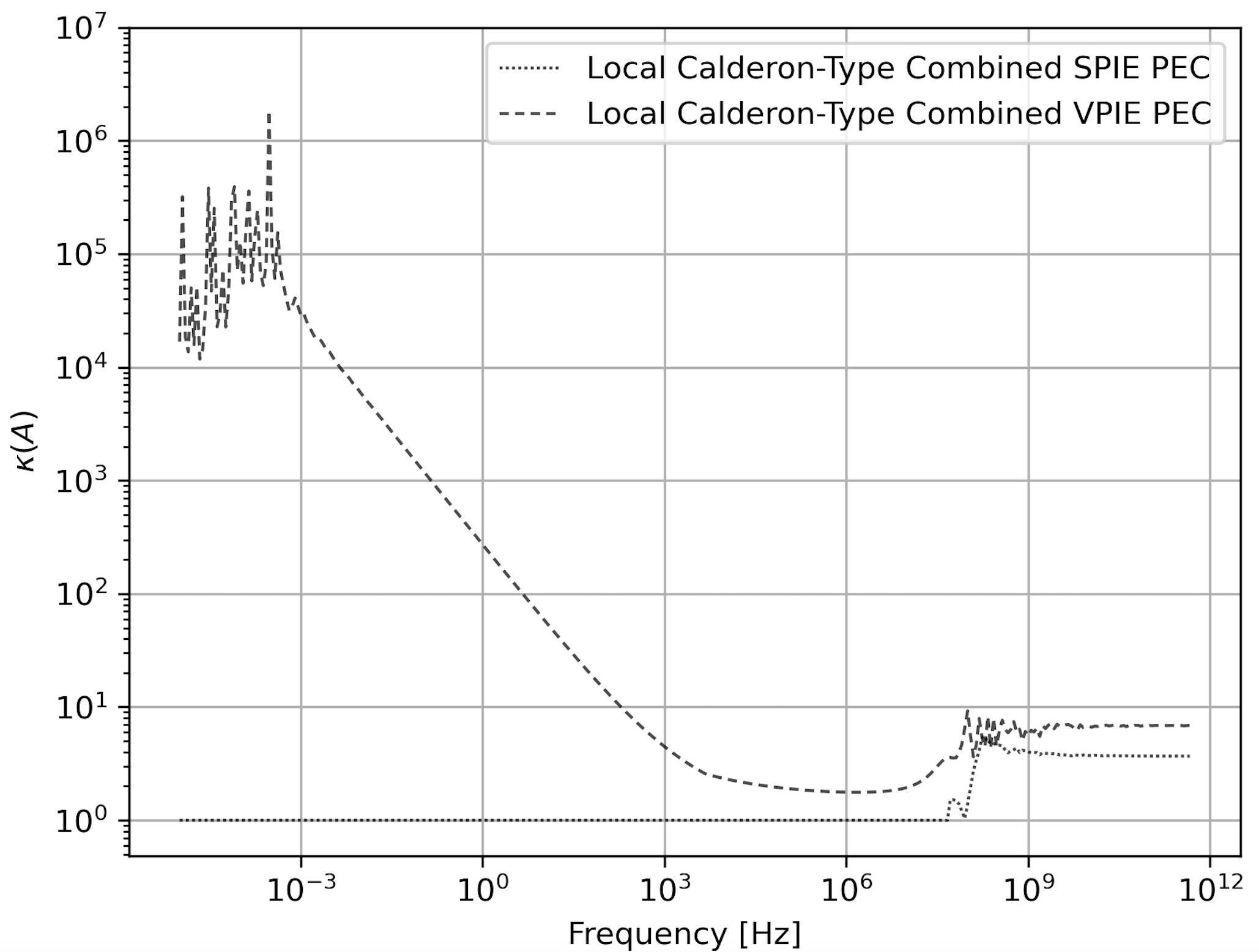}
\centering
\centering\caption{Plot of singular value conditioning.}
\label{fig:SVDCond}
\end{figure}

Finally, Fig. \ref{fig:SVDCond} is a singular value condition number plot of the LC-CVPIE and LC-CSPIE for frequencies ranging from 1e-5 Hz to 1e12 Hz. On the one hand, the singular value conditioning for the LC-CVPIE formulation increases at low frequencies. Due to the complex wavenumber, the upper right operator of \eqref{eq:CalderonPrecond} is not equal to zero
\begin{equation}
\lim_{\omega \to 0} ~\Tilde{\mathcal{L}^{t}}\mathcal{P}^{(2)} + \Tilde{\mathcal{Q}}^{(1)}\mathcal{S} \neq 0.
\end{equation}
On the other hand, the LC-CSPIE possesses flat, wideband singular value conditioning.  Fortunately, LC-CVPIE quickly converges using QMR, GMRES, and LGMRES to the Mie RCS at low frequency as is demonstrated in the next section. In other words, there is no practical cost to LC-CVPIE at low frequency.

\subsection{Analysis of Spheres Using Piece-Wise Basis}

\begin{figure}[h]
\includegraphics[width=8cm]{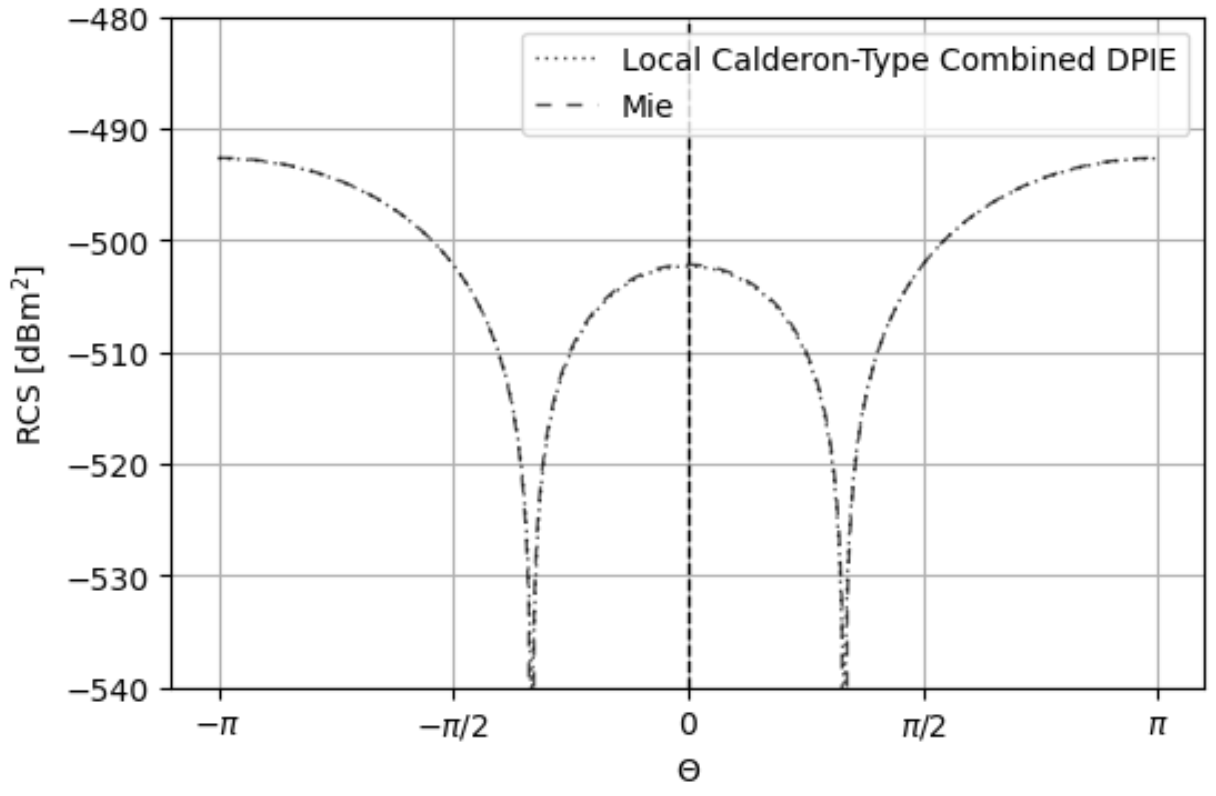}
\centering
\caption{RCS of a PEC sphere of unit radius.}
\label{fig:RCSLF_LC-CDPIE}
\end{figure}

Next, we predict RCS and benchmark the performance of LC-CVPIE, LC-CSPIE, and LC-CDPIE against the CVPIE, CSPIE, and CFIE formulations using Rao-Wilton-Glisson (RWG) functions for vector quantities and hat functions for scalar quantities. Figure \ref{fig:RCSLF_LC-CDPIE} shows the RCS of a PEC sphere with radius 1$m$ illuminated at 10$\mu$ Hz with a planewave of $\hat{\boldsymbol{z}}$ propagation axis, $\hat{\boldsymbol{x}}$ polarization, and the sphere is discretized with a mean edge length of $\lambda$/1e14. The LC-CVPIE and LC-CSPIE converged in 4 and 3 iterations, respectively, using QMR with a tolerance of 1e-5 and Method 1 to enforce the ZMC. The predicted RCS using LC-CDPIE solution agrees with the predicted RCS of the Mie solution. 

\begin{figure}[h]
\includegraphics[width=8cm]{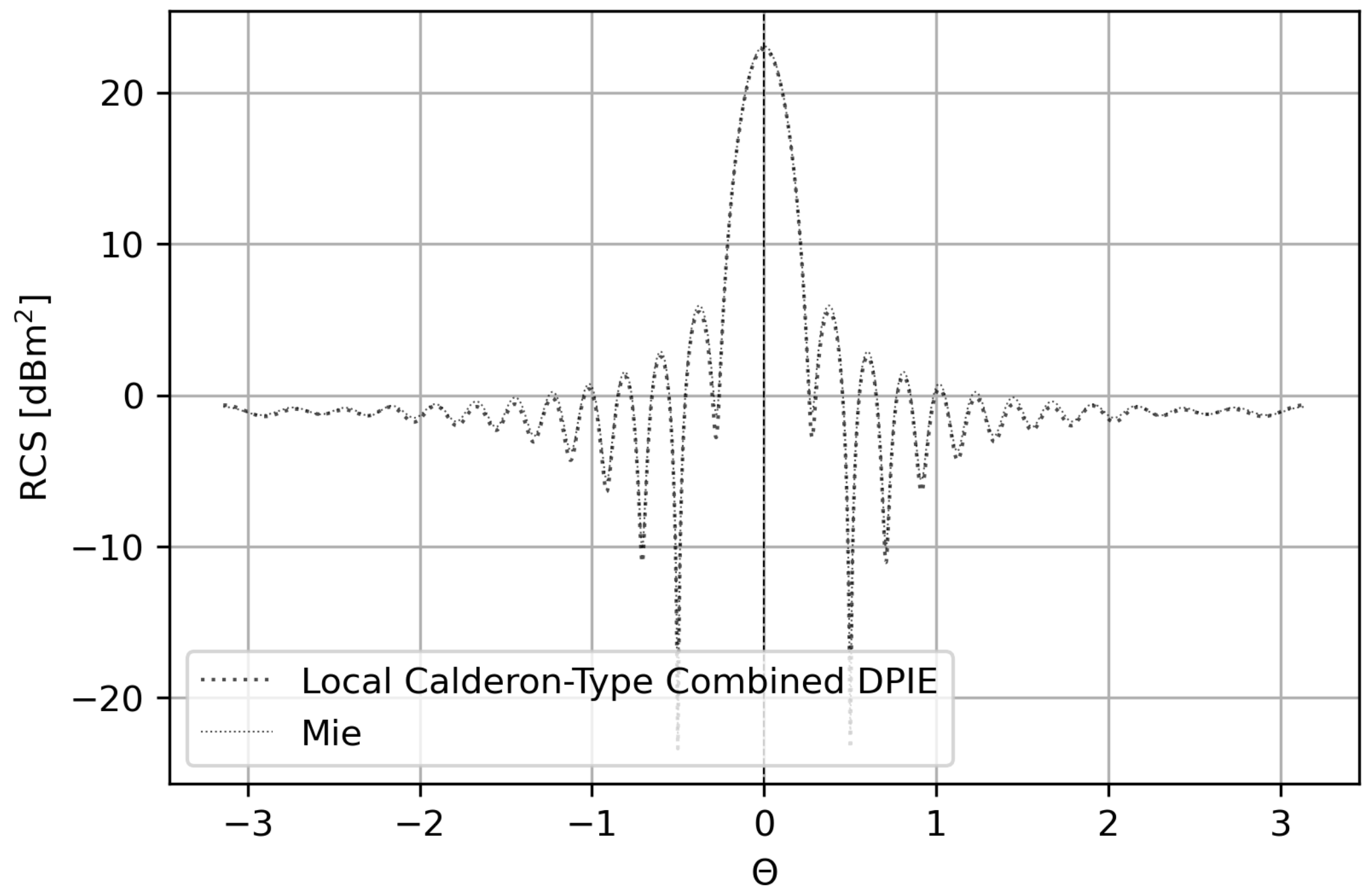}
\centering
\caption{RCS of a PEC sphere of radius 0.5 meters.} 
\label{Fig:HFRCS_LC-CDPIE}
\end{figure}

Furthermore, the LC-CDPIE formulation was tested with a PEC sphere of radius 0.5 meters excited by a 1500 MHz frequency planewave with $\hat{\boldsymbol{z}}$ propogation axis and $\hat{\boldsymbol{x}}$ polarization. The sphere discretization is $\lambda/10$, and the electrical length is $5\lambda$.  The MOM system was solved using QMR with a tolerance of 1e-12 and ZMC Method 1. The LC-CVPIE converged in 32 iterations while the LC-CSPIE converged in 24 iterations. The RCS predicted by the LC-CDPIE agrees with that of the Mie solution in Fig. \ref{Fig:HFRCS_LC-CDPIE}.

\begin{figure}[h]
\includegraphics[width=8cm]{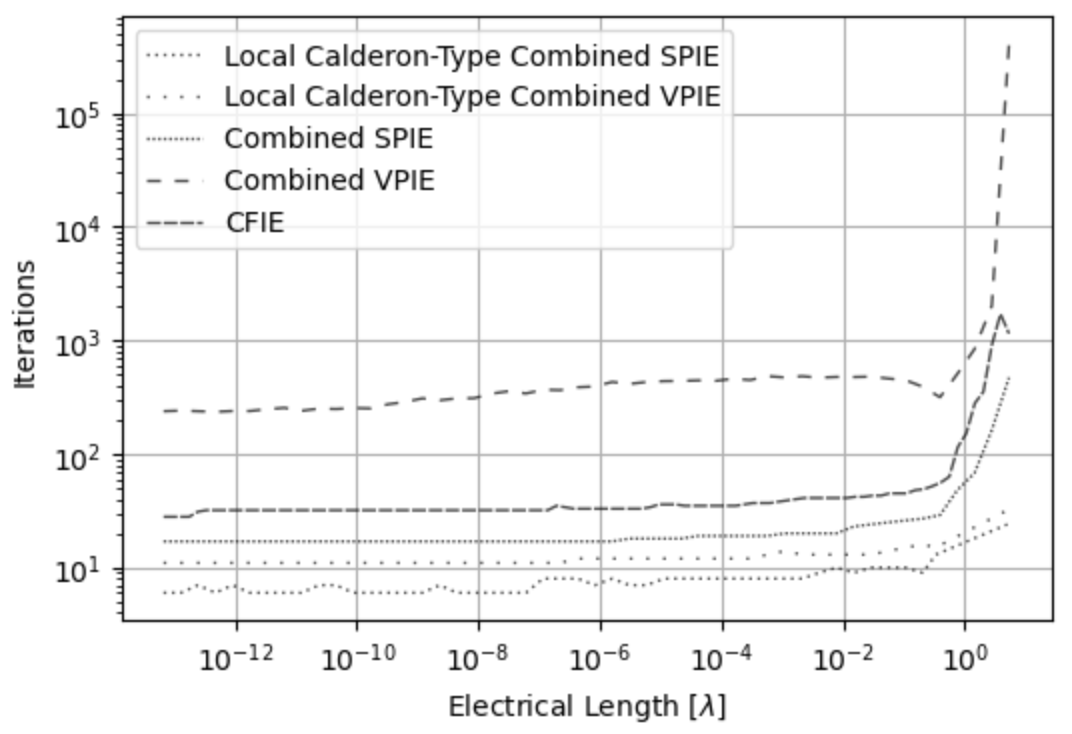}
\centering
\caption{Number of iterations for 1e-12 convergence using QMR for spheres of unit radius.} 
\label{Fig:iterFormulations}
\end{figure}

Next, the iteration count of each formulation to converge using QMR with a tolerance of 1e-12 and ZMC Method 1 for spheres whose electrical lengths range from 1e-12$\lambda$ to 5$\lambda$ is shown in Fig. \ref{Fig:iterFormulations}. Four different unit spheres are used in this test. For frequencies below 100 MHz, a sphere of average edge wavelength $\lambda/10$ at 100 MHz is used. For frequencies between 100 and 200 MHz, a sphere of average wavelength $\lambda/10$ at 200 Mhz is used. For frequencies between 200 and 400 MHz, a sphere of average wavelength $\lambda/10$ at 400 MHz is used. Finally, for frequencies between 400 and 750 MHz, a sphere of average wavelength $\lambda/10$ at 750 MHz is used. The LC-CVPIE and LC-CSPIE converge in less iterations than the CFIE, CVPIE, and CSPIE for all tested electrical lengths.  

The conditioning of the forward matrices in the MOM systems for a collection of spheres (whose discretization follows that which is described for Fig. \ref{Fig:HFRCS_LC-CDPIE}) of  radius 1$m$ using the LC-CVPIE and LC-CSPIE are recorded in Table 1 and 2. The ZMC methods (Method 1 or 2) are specified because they affect the conditioning of the forward matrix. For example, Lagrange multiplier in Method 2 increases the conditioning at higher frequencies while the Lagrange multiplier removes an eigenvalue near the origin of the complex plane at low frequencies. The formulation has constant eigenvalue conditioning for the recorded frequency range, but the singular value conditioning increases as the frequency lowers for reasons discussed above. However, large low-frequency singular value condition numbers for the LC-CVPIE have no practical impact upon the convergence or predicted RCS; the LC-CVPIE quickly converges in 4 iterations using QMR with a tolerance of 1e-5 at 10 $\mu$Hz to the RCS of the Mie solution for a unit sphere.

\begin{table}[h!]
\centering
\begin{tabular}{||c c c c||} 
 \hline
 ZMC & Freq. (Hz) & Sing. Val. Cond. & Eig. Cond. \\ [0.5ex] 
 \hline\hline
 Method 2 & 800e6 & 20.5 & 15.6 \\
 Method 2 & 600e6 & 19.2 & 15.1 \\ 
 Method 2 & 400e6 & 12.7 & 10.2 \\
 Method 2 & 200e6 & 9.0 & 6.5 \\
 Method 1 & 100e6 & 7.3 & 6.2 \\
 Method 1 & 1e5 & 7.1 & 5.9 \\
 Method 1 & 1e4 & 135 & 5.9 \\
 Method 1 & 1e3 & 1.3e4 & 5.9 \\
 Method 1 & 1e2 & 1.3e6 & 5.9 \\
 Method 1 & 1e1 & 1.3e9 & 5.9 \\
 Method 1 & 1e0 & 1.3e12 & 5.9 \\
 Method 1 & 1e-1 & 1.3e15 & 5.9 \\
 Method 1 & 1e-2 & 1.3e16 & 5.9 \\
 Method 1 & 1e-3 & 1.3e18 & 5.9 \\
 Method 1 & 1e-4 & 1.3e20 & 5.9 \\
 Method 1 & 1e-5 & 1.3e22 & 5.9 \\[1ex]
 \hline
\end{tabular}
\vspace{2mm}
\caption{Local Calder\'{o}n-Type Combined VPIE}
\label{table:1}
\end{table}

\begin{table}[h!]
\centering
\begin{tabular}{||c c c c||} 
 \hline
 ZMC & Freq. (Hz) & Sing. Val. Cond. & Eig. Cond. \\ [0.5ex] 
 \hline\hline
 Method 2 & 800e6 & 7.2 & 6.6 \\
 Method 2 & 600e6 & 6.2 & 6 \\ 
 Method 2 & 400e6 & 5.2 & 5.2 \\
 Method 2 & 200e6 & 6.5 & 6.4 \\
 Method 1 & 100e6 & 5.4 & 5.4 \\
 Method 1 & 1e5 & 5.7 & 5.7 \\
 Method 1 & 1e4 & 5.7 & 5.7 \\
 Method 1 & 1e3 & 5.7 & 5.7 \\
 Method 1 & 1e2 & 5.7 & 5.7 \\
 Method 1 & 1e1 & 5.7 & 5.7 \\
 Method 1 & 1e0 & 5.7 & 5.7 \\
 Method 1 & 1e-1 & 5.7 & 5.7 \\
 Method 1 & 1e-2 & 5.7 & 5.7 \\
 Method 1 & 1e-3 & 5.7 & 5.7 \\
 Method 1 & 1e-4 & 5.7 & 5.7 \\
 Method 1 & 1e-5 & 5.7 & 5.7 \\[1ex]
 \hline
\end{tabular}
\vspace{2mm}
\caption{Local Calder\'{o}n-Type Combined SPIE}
\label{table:2}
\end{table}

The LC-CDPIE is also tested with a multi-scale sphere of unit radius illuminated with a 400 MHz planewave that propagates along the  $\hat{\boldsymbol{z}}$  axis and is polarized along $\hat{\boldsymbol{x}}$ polarization. The  electrical length of the mesh edges ranges from $\lambda/11$ to $\lambda/406$. The LC-VPIE and LC-SPIE formulations converge more quickly than the CFIE formulation using QMR, TFQMR, and GMRES. Converging to a residue of 1e-12 requires less than 100 iterations for the LC-CSPIE (Group 1 in Fig. \ref{Fig:ResiduePlot}) and less than 210 iterations for the suggested LC-CVPIE (Group 2 in Fig. \ref{Fig:ResiduePlot}) while the CFIE requires over 24,000 iterations for TFQMR as well as QMR and will not converge using GMRES (Group 3 in Fig. \ref{Fig:ResiduePlot}).  

\begin{figure}
\includegraphics[width=5cm]{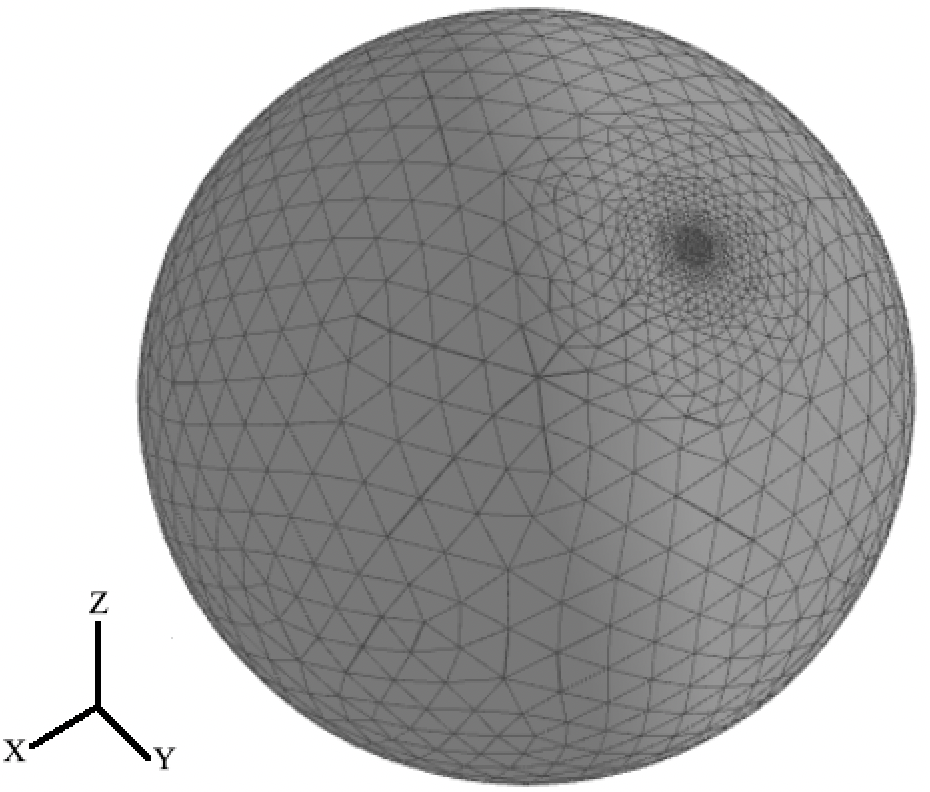}
\centering
\caption{Plot of multi-scale sphere.} 
\end{figure}

\begin{figure}
\includegraphics[width=8cm]{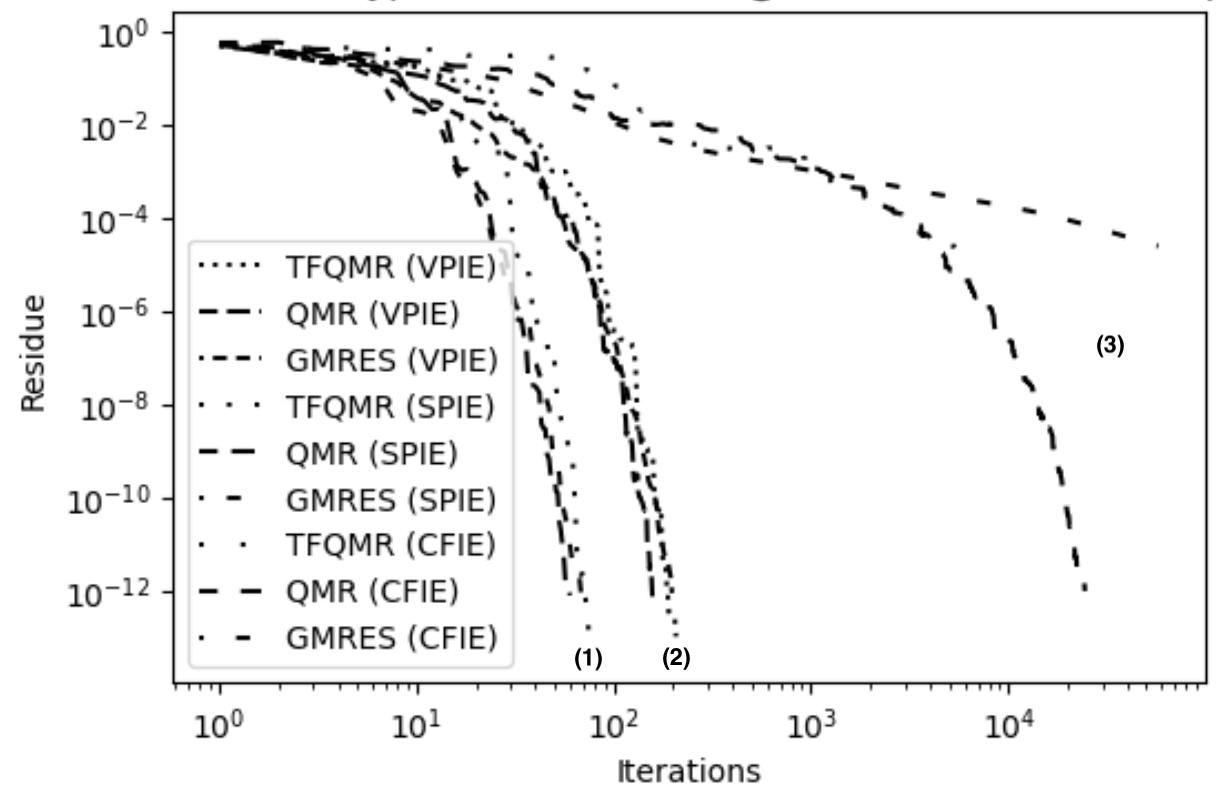}
\centering
\caption{Residue plot for multi-scale sphere.} 
\label{Fig:ResiduePlot}
\end{figure}

\subsection{Analysis of Non-Canonical Geometries Using Piece-Wise Basis Sets}

\begin{figure}
\includegraphics[width=8cm]{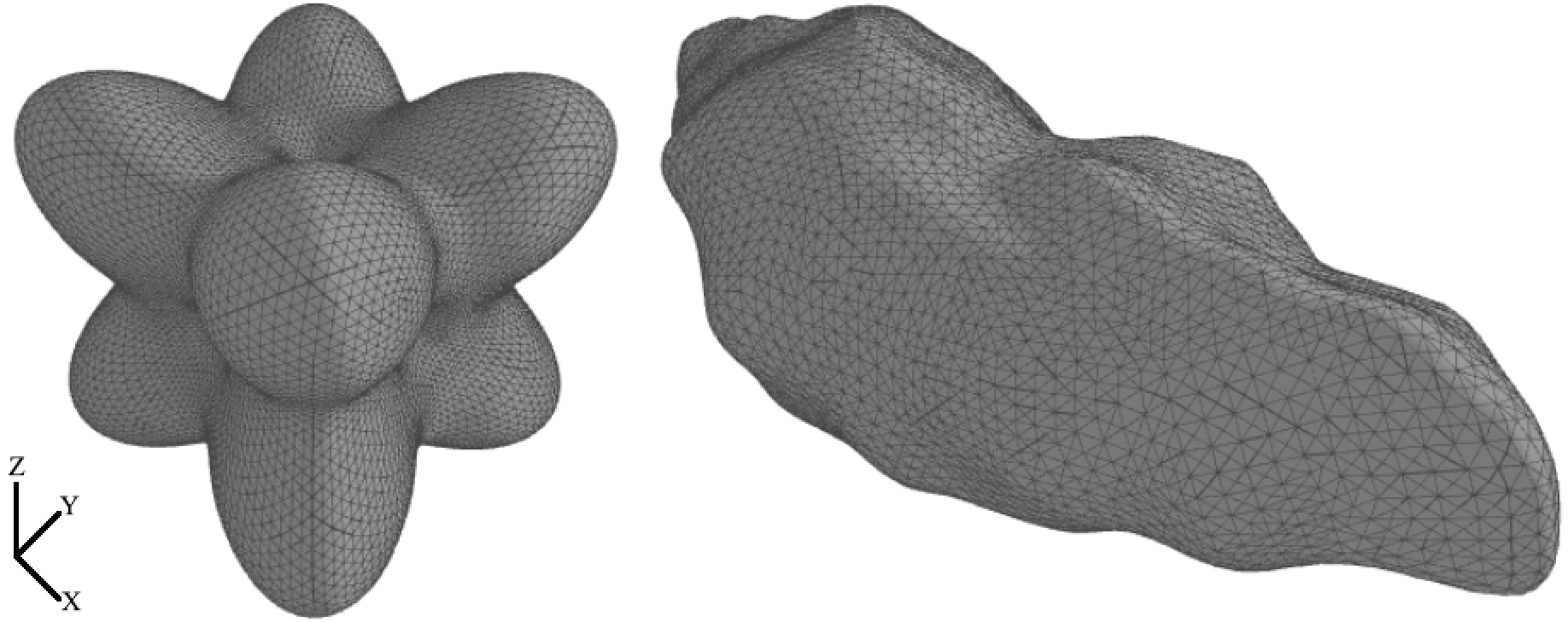}
\centering
\caption{Plot of bumpy cube and NASA Geographos asteroid geometries.} 
\label{Fig:PlotCubeAsteroid}
\end{figure}

\begin{figure}
\includegraphics[width=8cm]{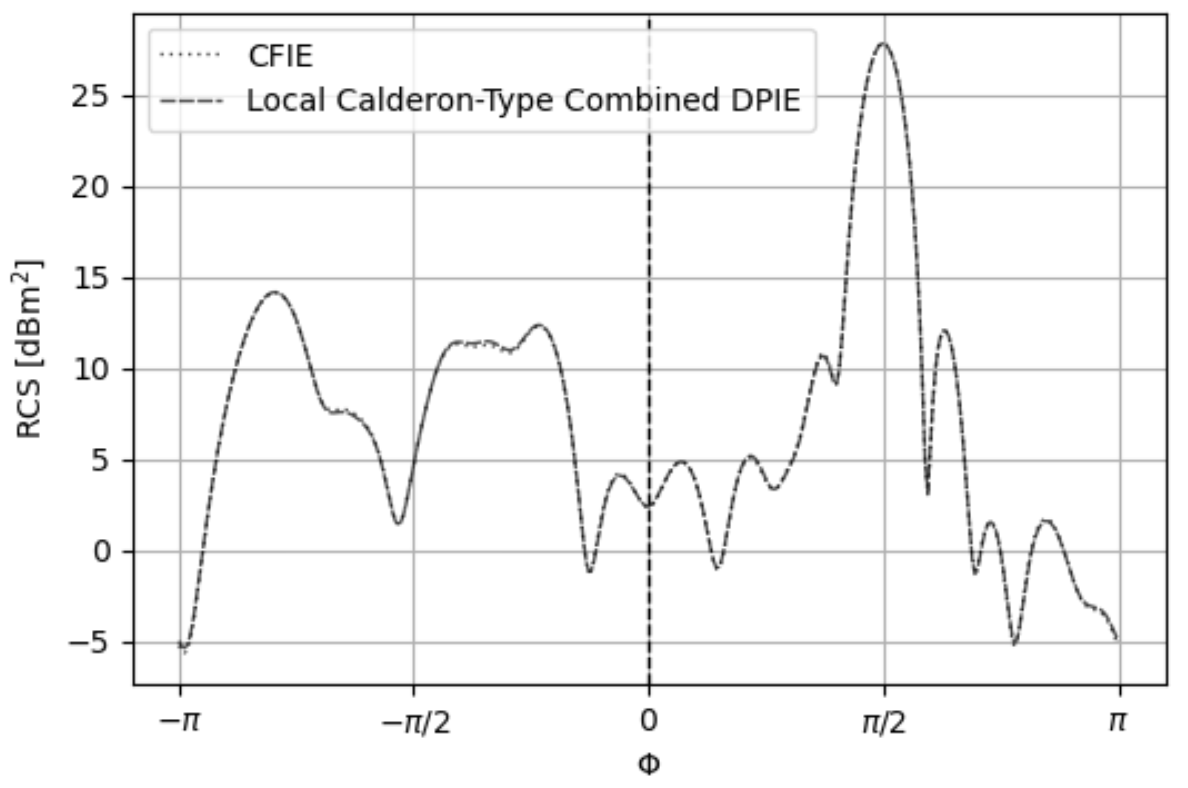}
\centering
\caption{RCS plot of the NASA Geographos asteroid.} 
\label{Fig:RCSAsteroid}
\end{figure}

\begin{figure}
\includegraphics[width=8cm]{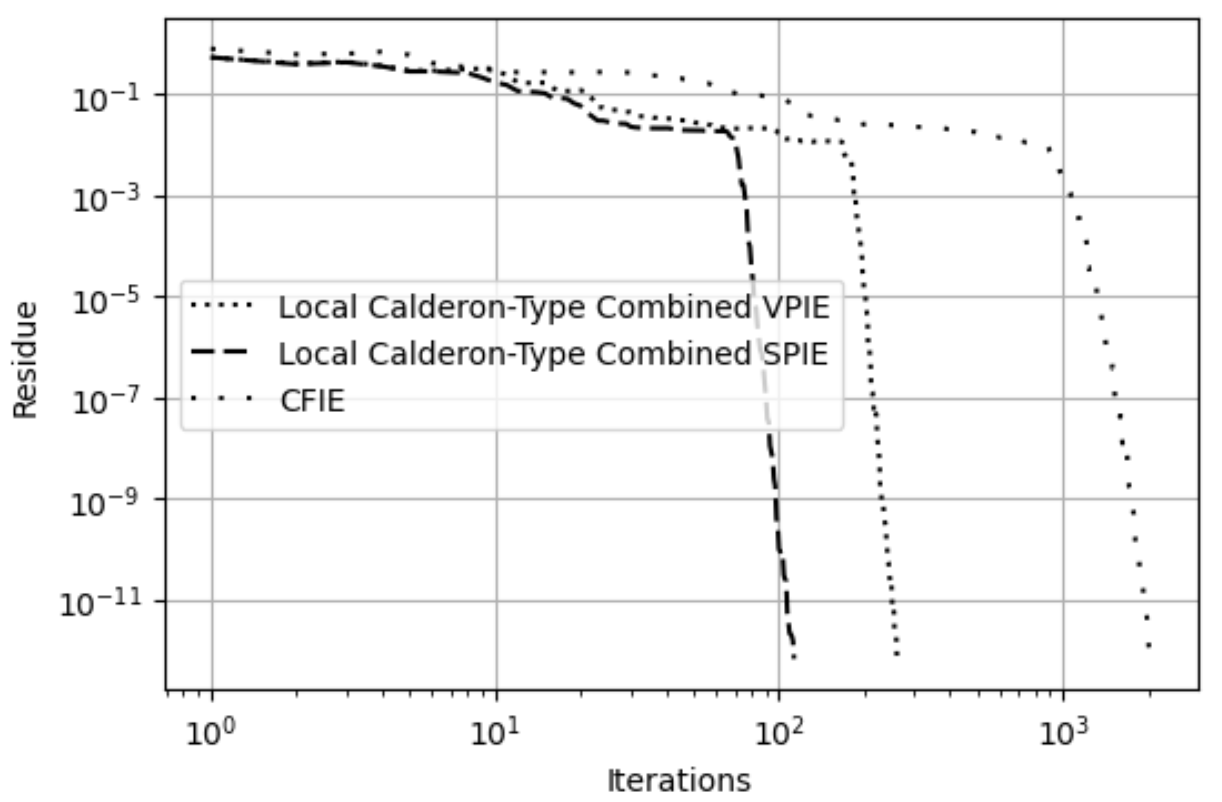}
\centering
\caption{Residue plot for the NASA Geographos asteroid.} 
\label{Fig:ResiduePlotAsteroid}
\end{figure}

The next set of benchmarking tests uses non-canonical geometries. We select a  bumpy cube,  a  NASA Geographos asteroid as shown in Fig. \ref{Fig:PlotCubeAsteroid} and a sharp pencil as shown in Fig. \ref{Fig:PlotPencil}. The NASA Geographos asteroid is illuminated by a 240 MHz planewave that propagates along $\hat{\boldsymbol{y}}$ and polarized along  $\hat{\boldsymbol{x}}$. The electrical lengths of the mesh edges range from $\lambda/13$ to $\lambda/64$, and the electrical length of the object is $4\lambda$. The MOM systems are solved using QMR with tolerance 1e-12, ZMC Method 1, and the predicted RCS of the LC-CDPIE and CFIE agree in Fig. \ref{Fig:RCSAsteroid}. The LC-CVPIE and LC-CSPIE formulations converge in 258 and 112 iterations, respectively, while the CFIE converges in 2027 iterations. The plot of residues and iteration counts in Fig. \ref{Fig:ResiduePlotAsteroid} shows the LC-CVPIE and LC-CSPIE formulations converge more quickly to an arbitrary residue than the CFIE for the asteroid.

\begin{figure}
\includegraphics[width=8cm]{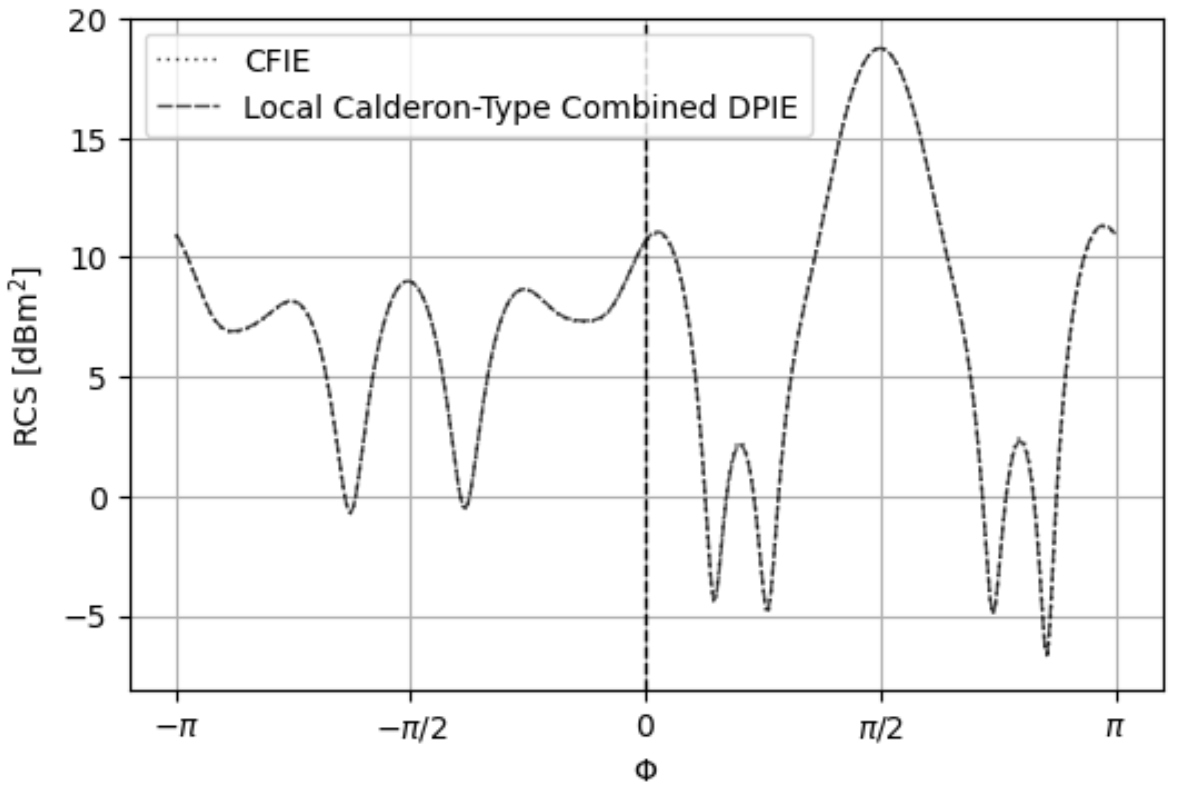}
\centering
\caption{RCS plot of a bumpy cube.} 
\label{Fig:RCSCube}
\end{figure}

\begin{figure}
\includegraphics[width=8cm]{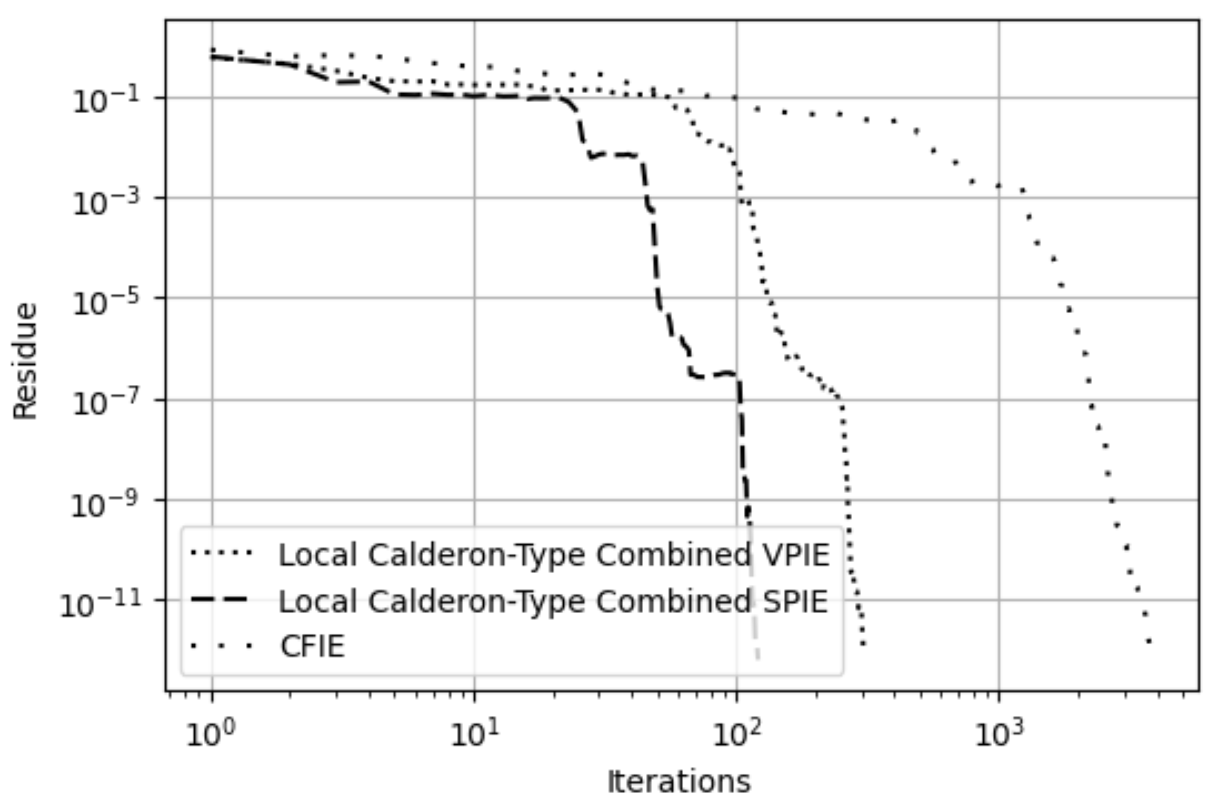}
\centering
\caption{Residue plot for a bumpy cube.} 
\label{Fig:ResiduePlotCube}
\vspace{3mm}
\includegraphics[width=8cm]{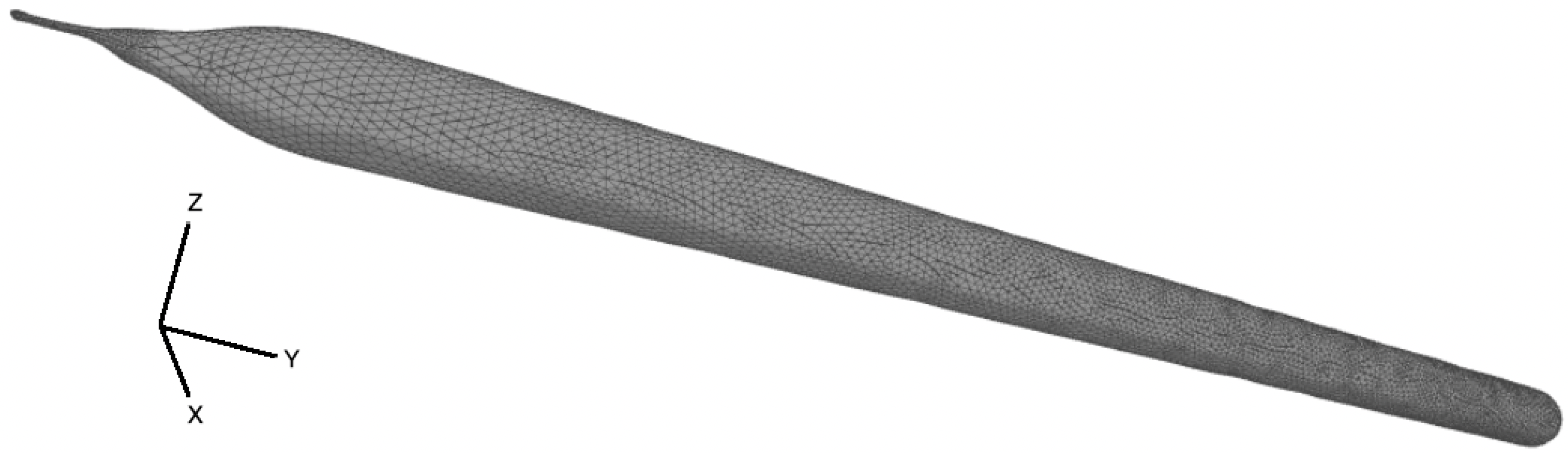}
\centering
\caption{Plot of sharp pencil geometry.} 
\label{Fig:PlotPencil}
\end{figure}

The bumpy cube is illuminated by a 400 MHz planewave propagating along  $\hat{\boldsymbol{y}}$   and polarized along $\hat{\boldsymbol{x}}$. The electrical lengths of the mesh edges for the bumpy cube range from $\lambda/12$ to $\lambda/50$. Furthermore, the object fits in a cube of size $1.67\lambda$. Again, the MOM systems are solved using QMR with a tolerance of 1e-12, ZMC Method 1, and the predicted RCS of the LC-CDPIE and CFIE agree in Fig. \ref{Fig:RCSCube}. The LC-CVPIE and LC-CSPIE formulations converge in 305 and 120 iterations, respectively, while the CFIE converges in 3761 iterations. Also, the plot of residues and iteration counts in Fig. \ref{Fig:ResiduePlotCube} shows the LC-CVPIE and LC-CSPIE formulations converge more quickly to an arbitrary residue than the CFIE for the bumpy cube as well as the asteroid. 

\begin{figure}
\includegraphics[width=8cm]{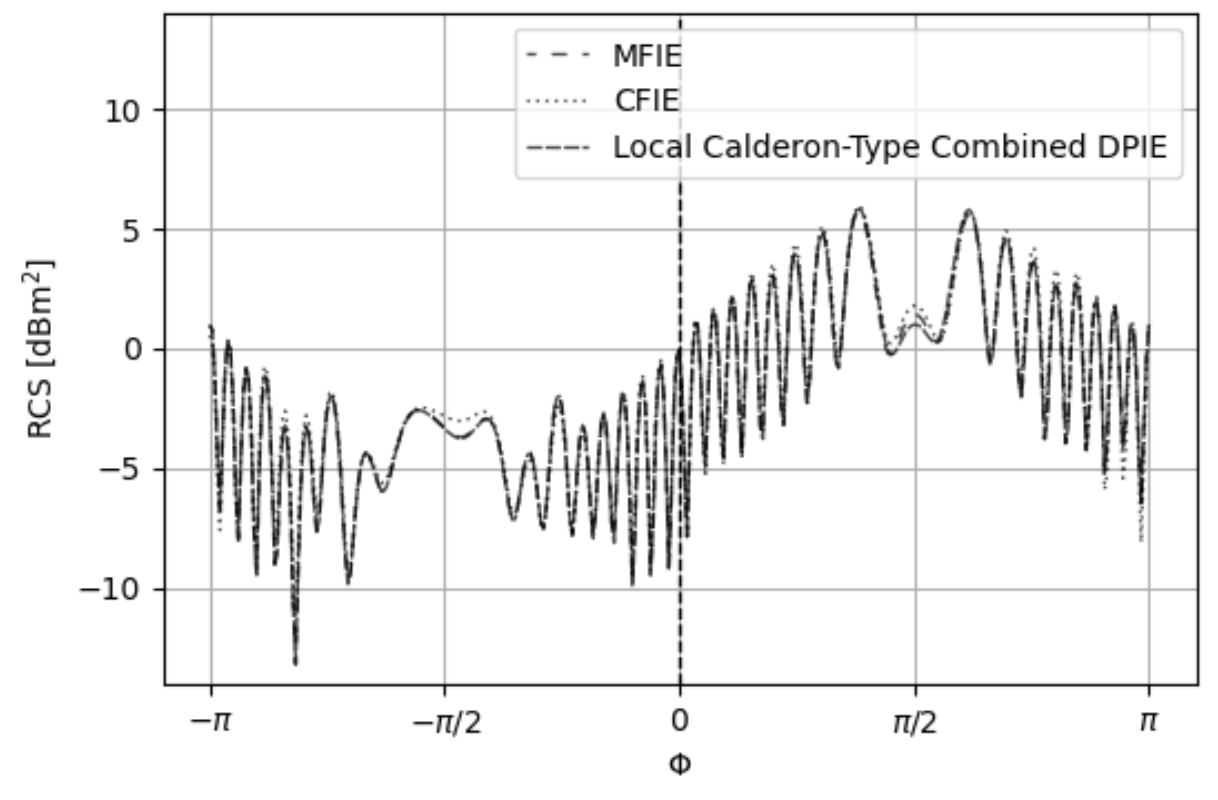}
\centering
\caption{RCS plot for a sharp pencil.} 
\label{Fig:RCSPencil}
\end{figure}

\begin{figure}
\includegraphics[width=8cm]{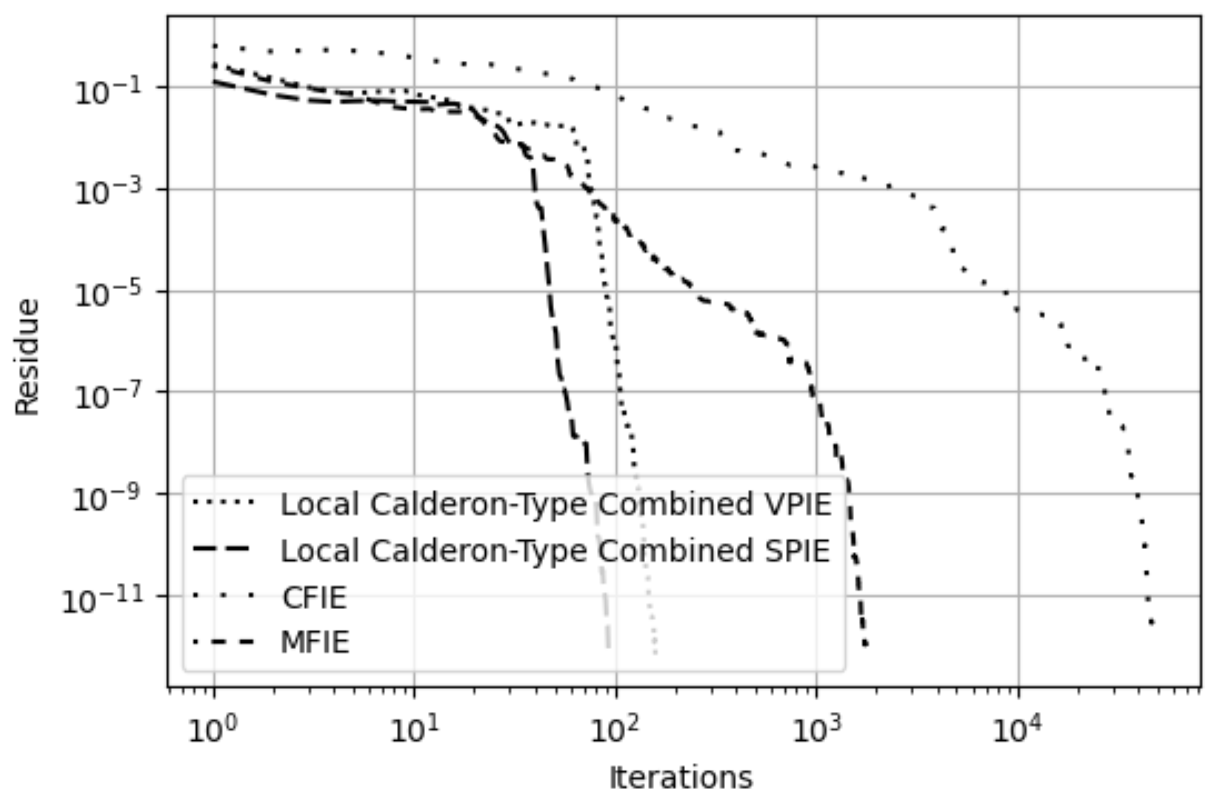}
\centering
\caption{Residue plot for a sharp pencil.} 
\label{Fig:ResiduePlotPencil}
\end{figure}

The last and most challenging test is analysis of scattering from a sharp pencil as shown in Fig. \ref{Fig:PlotPencil}. This geometry's curvature max-mean curvature is 2585 due to the sharp, narrow tip. The sharp pencil is illuminated by a 269 MHz planewave with $\hat{\boldsymbol{y}}$ propagation axis and $\hat{\boldsymbol{x}}$ polarization. The electrical lengths of the mesh edges range from $\lambda/13$ to $\lambda/5648$. This range is greater than the multi-scale sphere, bumpy cube, and asteroid. Also, the pencil's electrical length is $9\lambda$. The MOM systems are solved using QMR with a tolerance of 1e-12 and ZMC Method 1. The LC-CDPIE, CFIE, and MFIE RCS plots agree in Fig. \ref{Fig:RCSPencil}, but the LC-CDPIE and MFIE agree most closely while the CFIE departs from both near $\phi = -\pi/2~\text{and}~\pi/2$. The LC-CVPIE and LC-CSPIE formulations converge in 159 and 90 iterations, respectively, while the CFIE converges in 47045 iterations and the MFIE in 1761 iterations. If the fine-grain localization approach is not used but rather the global max-mean curvature localization approach discussed in Section \ref{Sec:Localization} is used, the LC-CVPIE and LC-CSPIE converge in 34764 and 91 iterations, respectively. Again, the plot of residues and iteration counts in Fig. \ref{Fig:ResiduePlotPencil} show the LC-CVPIE and LC-CSPIE formulations converge more quickly to any residue than both the CFIE and MFIE.

\vspace{2mm}

\vspace{2mm}

\vspace{2mm}

\vspace{2mm}

\newpage

\section{Conclusion}

This paper has presented new Calder\'{o}n-type identities and used these identities to construct the LC-CSPIE, LC-CVPIE, and LC-CDPIE formulations for solving the PEC scattering problem with RWG and hat functions. The spectra of these formulations were shown to be bounded and non-zero, and the increase in singular value conditioning at low-frequency for the LC-CVPIE was shown to be of no practical importance for iterative solvers. Indeed, these novel formulations are well-suited for efficiently predicting the scattering from multi-scale, real-world PEC geometries.


%

\section*{Acknowledgment}
This work was supported in part by the National Science Foundation under Grant No. CCF-1822932. This research used resources of Michigan State University's High Performance Computing Center (MSU HPCC).

\appendices

\section{Definition of Operators}

\begin{subequations}
    \begin{align}
        \mathcal{S}\begin{pmatrix}
            \textbf{x} \\
            \text{x} \\
        \end{pmatrix} = & \int G(\textbf{r}, \textbf{r}')\begin{pmatrix}
            \textbf{x} \\
            \text{x} \\
        \end{pmatrix} dS' \\
        \mathcal{D}\left( x \right) = & -\nabla\cdot\mathcal{S}\left( \hat{\textbf{n}} ~ x \right) \\
        \mathcal{N} \left( x \right) = & -\hat{\textbf{n}}\cdot\nabla\nabla\cdot\mathcal{S}\left( \hat{\textbf{n}} ~ x \right) \\
        \mathcal{D}' \left( x \right) = & \hat{\textbf{n}}\cdot\nabla\mathcal{S}\left( x \right) \\
        \mathcal{K}' \left( \textbf{x} \right) = & - \nabla\times\mathcal{S}\left(\hat{\textbf{n}}\times\textbf{x} \right) \\
        \mathcal{J}^{(2)}\left( \textbf{x} \right) = & \hat{\textbf{n}}\times\mathcal{S}\left( \hat{\textbf{n}}\times\textbf{x} \right) \\
        \mathcal{J}^{(3)}\left( \textbf{x} \right) = & \hat{\textbf{n}}\cdot\mathcal{S}\left(\hat{\textbf{n}}\times\textbf{x} \right) \\
        \mathcal{J}^{(4)}\left( \textbf{x} \right) = & \nabla\cdot\mathcal{S}\left(\hat{\textbf{n}}\times\textbf{x}\right) \\
        \mathcal{L}\left( \textbf{x} \right) = & \frac{1}{\kappa^2}\nabla\times\nabla\times\mathcal{S}\left( \textbf{x} \right) \\
        \mathcal{K}\left( \textbf{x} \right) = & \hat{\textbf{n}}\times\nabla\times\mathcal{S}\left( \textbf{x} \right) \\
        \mathcal{M}^{(3)}\left( \textbf{x} \right) = & \hat{\textbf{n}}\cdot\nabla\times\mathcal{S}\left( \textbf{x} \right) \\
        \mathcal{P}^{(2)}\left( \textbf{x} \right) = & \hat{\textbf{n}}\times\nabla\mathcal{S}\left( x \right) \\
        \mathcal{Q}^{(1)}\left( x \right) = & \hat{\textbf{n}}\times\hat{\textbf{n}}\times\nabla\times\mathcal{S}\left(\hat{\textbf{n}} ~ x \right) \\
        \mathcal{Q}^{(2)}\left( x \right) = & \hat{\textbf{n}}\times\mathcal{S}\left( \hat{\textbf{n}} ~ x \right) \\
        \mathcal{Q}^{(3)} \left( x \right) = & \hat{\textbf{n}}\cdot\mathcal{S}\left( \hat{\textbf{n}} ~ x \right)
    \end{align}
\end{subequations}

Prime denotes an adjoint operator and superscript t is shorthand for $\mathcal{O}^t = \hat{\textbf{n}}\times\hat{\textbf{n}}\times\mathcal{O}$.

\section{Spherical Harmonics}

\begin{subequations}
    \begin{align}
        Y^m_n \left( \textbf{r} \right) = & \sqrt{\frac{2n + 1}{4\pi}\frac{\left( n - m \right)!}{\left( n + m \right)!}} P^m_n \left( \cos{\theta}\right)e^{jm\phi} \\
        \mathbf{\Psi}^m_n\left( \hat{\textbf{r}} \right) = & - \hat{\textbf{r}}\times\mathbf{\Phi}^m_n\left( \hat{\textbf{r}} \right) = c_n r\nabla Y^m_n \left( \hat{\textbf{r}} \right) \\
        \mathbf{\Phi}^m_n\left( \hat{\textbf{r}} \right) = & \hat{\textbf{r}}\times\mathbf{\Psi}^m_n\left( \hat{\textbf{r}} \right) = ~ c_n \hat{\textbf{r}}\times \nabla Y^m_n \left( \hat{\textbf{r}} \right) \\
        c_n = & \begin{cases}
            1 & n = 0 \\
            \frac{1}{\sqrt{n\left( n + 1 \right)}} & n \ne 0 \\
        \end{cases} \\
    \end{align}
\end{subequations}

with $P^m_n\left(\cos{\theta}\right)$ being the associated Legendre polynomials and $n \ge 0$ and $|m| \le n$.

\ifCLASSOPTIONcaptionsoff
  \newpage
\fi

\bibliography{paper}{}
\bibliographystyle{ieeetr}



%

    \begin{IEEEbiography}[{\includegraphics[width=1in,height=1.25in,clip,keepaspectratio]{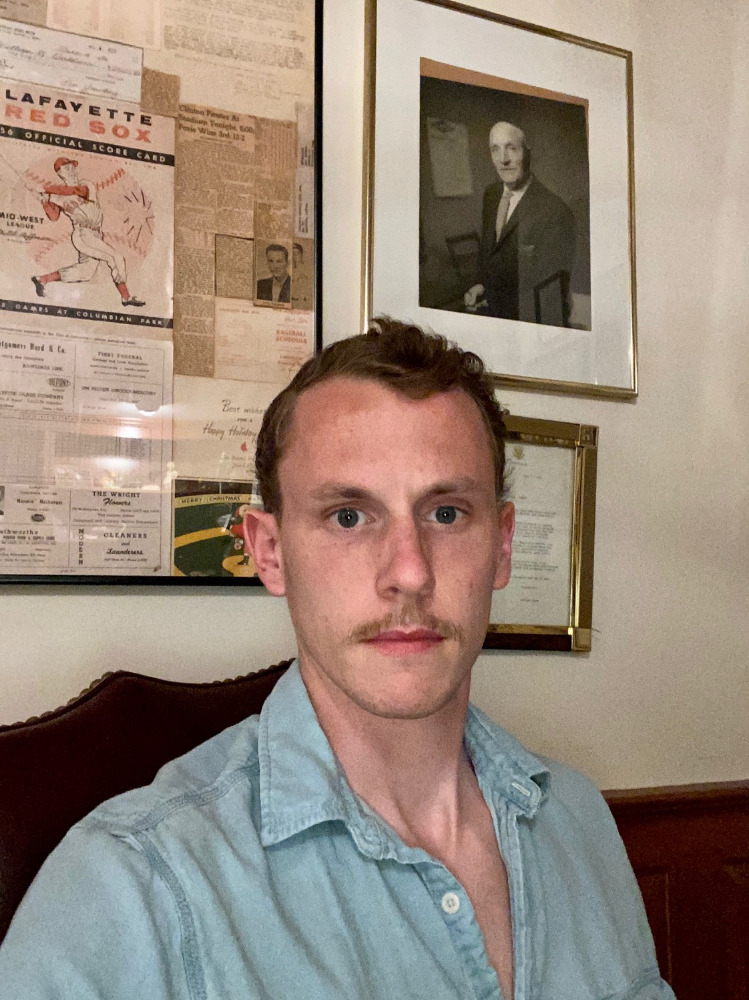}}]{Jacob Hawkins}
        (S'14) received the B.S. degree in physics from United States Air Force Academy, Colorado Springs, CO in 2014 and the M.S. degree in electrical engineering from Michigan State University in 2022 where he is currently a Ph.D. candidate.
        His research interests include well-conditioned formulations, surface integral equations, fast methods, and parallel computing for electrogmatics.
    \end{IEEEbiography}

    \begin{IEEEbiography}[{\includegraphics[width=1in,height=1.25in,clip,keepaspectratio]{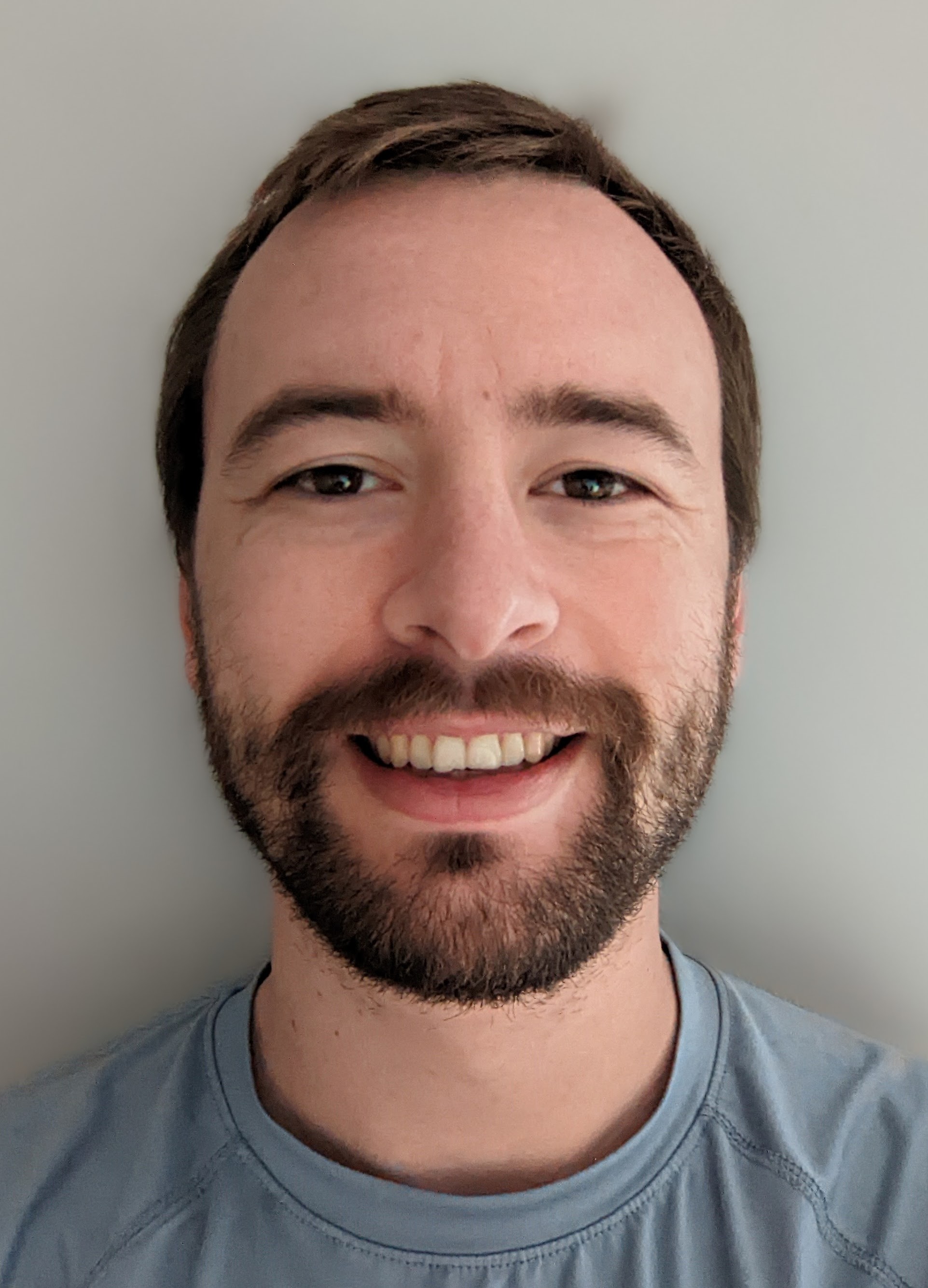}}]{Luke Baumann}
        (S'14) received the B.S. degree in computer engineering from California Polytechnic State University, San Luis Obsipo, CA in 2014 and the M.S. degree in electrical engineering from Michigan State University in 2021 where he is currently a Ph.D. candidate.
        His research interests include well-conditioned formulations, surface integral equation methods, and fast algorithms in electromagnetics.
    \end{IEEEbiography}

    \begin{IEEEbiography}[{\includegraphics[width=1in,height=1.25in,clip,keepaspectratio]{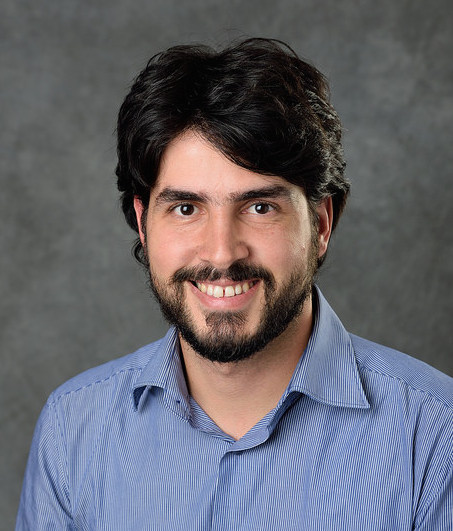}}]{H. M. Aktulga} 
        is an Associate Professor in the Department of Computer Science and Engineering at Michigan State University (MSU) where he directs the Scalable Parallel Technologies and Algorithms (SParTA) group. He primarily works on the design and development of parallel algorithms, numerical methods, performance models and software systems to harness the full potential of state-of-the-art HPC platforms for challenging problems in large scale scientific computing and big-data analytics. Dr. Aktulga has been an author/co-author in over 50 research articles in prestigious journals, conferences and book series. He is a recipient of the 2019 NSF CAREER award.
	\end{IEEEbiography}
	
	\begin{IEEEbiography}[{\includegraphics[width=1in,height=1.25in,clip,keepaspectratio]{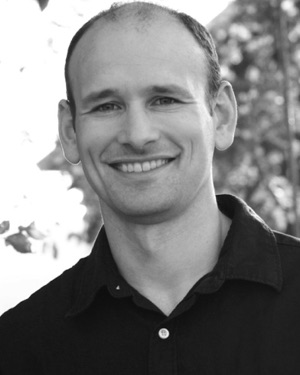}}]{D. Dault} 
        (M'17, SM'22) received the B.S. degree in electrical engineering and the Ph.D. degree in computational electromagnetics from Michigan State University, East Lansing, MI in 2010 and 2015, respectively. After graduation he worked as a computational electromagnetics engineer at the Riverside Research Institute in 2015, and in 2015 joined the US Air Force Research Laboratory, Wright-Patterson Air Force Base in Dayton OH where he researches and develops high performance computational electromagnetic tools. Dr. Dault was a recipient of the US Department of Defense National Defense in Science and Engineering Graduate Fellowship in 2010 and the National Science Foundation Graduate Research Fellowship in 2011.  In 2017 Dr. Dault was part of a team that won the National Defense Industrial Association Lt Gen Thomas R. Ferguson, Jr. Systems Engineering Excellence Team Award.  In 2022 Dr. Dault received the Air Force Research Laboratory Early Career Award for contributions in computational electromagnetics.
	\end{IEEEbiography}
	
	\begin{IEEEbiography}[{\includegraphics[width=1in,height=1.25in,clip,keepaspectratio]{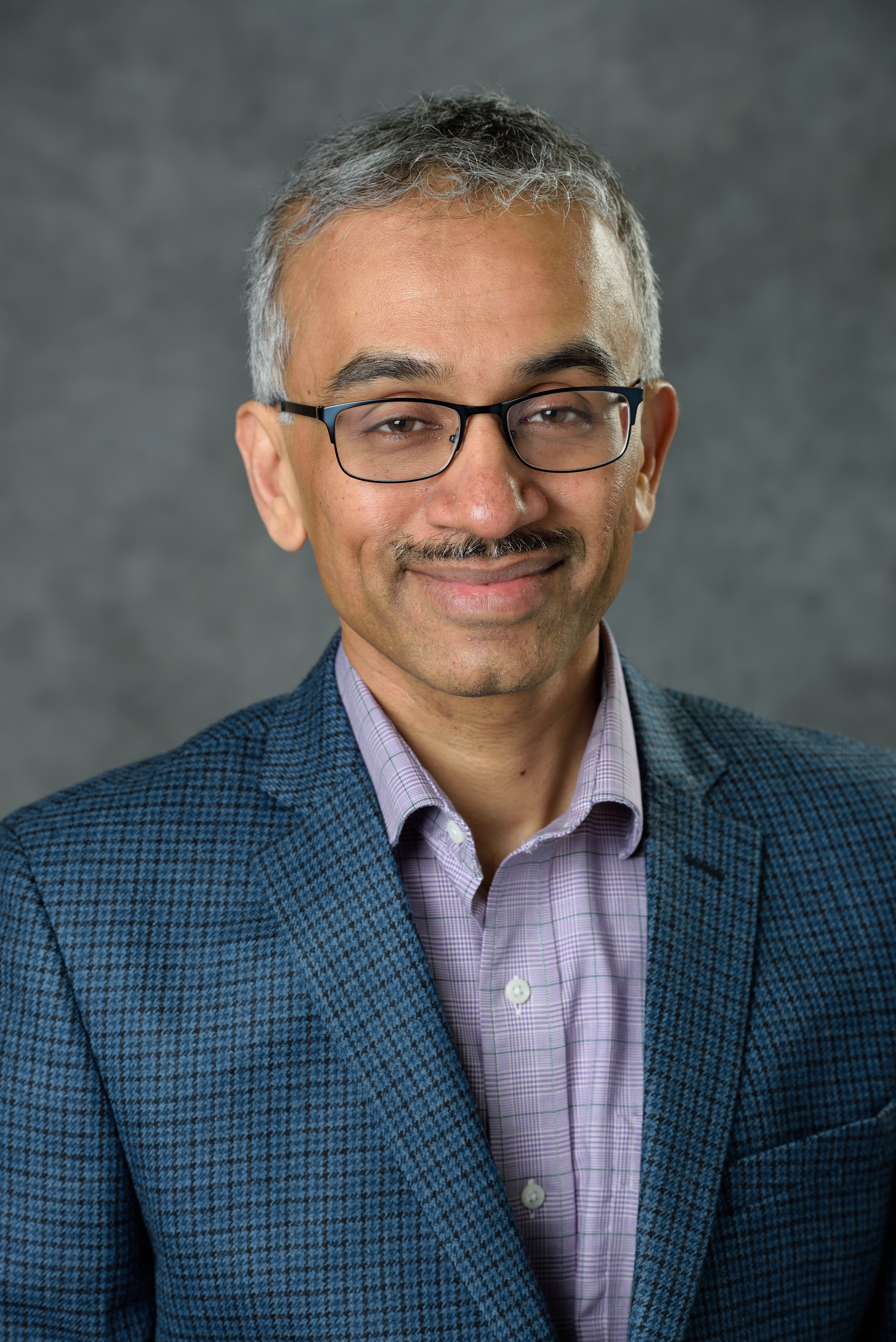}}]{B. Shanker}
		received his B'Tech from the Indian Institute of Technology, Madras, India in 1989, M.S. and Ph.D in 1992 and 1993, respectively, from The Pennsylvania State University. From 1993 to 1996 he was a research associate in the Department of Biochemistry and Biophysics at Iowa State University where he worked on the Molecular Theory of Optical Activity. From 1996 to 1999 he was with the Center for Computational Electromagnetics at the University of Illinois at Urbana-Champaign as a Visiting Assistant Professor, and from 1999-2002 with the Department of Electrical and Computer Engineering at Iowa State University as an Assistant Professor. From 2017, he was a University Distinguished Professor (an honor accorded to about 2\% of tenure system MSU faculty members) in the Department of Electrical and Computer Engineering at Michigan State University, and the Department of Physics and Astronomy. Currently, he is a Professor and Chair of Electrical and Computer Engineering at The Ohio State University. At Michigan State University, he was appointed Associate Chair of the Department of Computational Mathematics, Science and Engineering, a new department at MSU and was a key player in building this Department. Earlier he served as the Associate Chair for Graduate Studies in the Department of Electrical and Computer Engineering from 2012-2015, and the Associate Chair for Research in ECE from 2019-2022. He has authored/co-authored around 450 journal and conference papers and presented a number of invited talks. His research interests include all aspects of computational electromagnetics (frequency and time domain integral equation based methods, multi-scale fast multipole methods, fast transient methods, higher order finite element and integral equation methods), propagation in complex media, mesoscale electromagnetics, and particle and molecular dynamics as applied to multiphysics and multiscale problems. He was an Associate Editor for IEEE Antennas and Wireless Propagation Letters (AWPL), IEEE Transactions on Antennas and Propagation, and Topical Editor for Journal of Optical Society of America: A. He is a full member of the USNC-URSI Commission B. He is Fellow of IEEE (class 2010), elected for his contributions to time and frequency domain computational electromagnetics. He has also been awarded the Withrow Distinguished Junior scholar (in 2003), Withrow Distinguished Senior scholar (in 2010), the Withrow teaching award (in 2007), and the Beal Outstanding Faculty award (2014).
	\end{IEEEbiography}




\end{document}